\documentclass[14pt, preprint2]{aastex62}
\graphicspath{{./}{Plots/}}
\usepackage{float}
\makeatletter
\let\newfloat\newfloat@ltx
\makeatother
\usepackage{graphicx, subfigure, algorithm, algorithmic, amsmath}
\usepackage[normalem]{ulem}
\newcommand{\project}[1]{\textsl{#1}}
\newcommand{\thecannon}{\project{The~Cannon}} 
 
\newcommand{\apogee}{\project{\textsc{apogee}}}

\newcommand{\aspcap}{\project{\textsc{aspcap}}}

\newcommand{\Gaia}{\project{Gaia}}
\newcommand{\Gaiaeso}{\project{Gaia--\textsc{eso}}}
\newcommand{\galah}{\project{\textsc{galah}}}

\newcommand{\code}[1]{\texttt{#1}}
 
\newcommand{\teff}{\mbox{$T_{\rm eff}$}}

\newcommand{\feh}{\mbox{$\rm [Fe/H]$}}
\newcommand{\mgfe}{\mbox{$\rm [Mg/Fe]$}}
\newcommand{\xfe}{\mbox{$\rm [X/Fe]$}}
\newcommand{\alphafe}{\mbox{$\rm [\alpha/Fe]$}}

\newcommand{\logg}{\mbox{$\log g$}}

\newcommand{\rgal}{\mbox{$R_{\text{GAL}}$}}

\newcommand{\rbirth}{$R_\text{birth}$}
\newcommand{\cefe}{\mbox{$\rm [Ce/Fe]$}}
\newcommand{\bafe}{\mbox{$\rm [Ba/Fe]$}}
\newcommand{\cfe}{\mbox{$\rm [C/Fe]$}}
\newcommand{\eufe}{\mbox{$\rm [Eu/Fe]$}}
\newcommand{\sigint}{$\sigma_\text{intrinsic}$}

\begin{document}

\title{The chemical enrichment of the Milky Way disk evaluated using conditional abundances}

\author{Bridget L. Ratcliffe}
\affil{Department of Statistics, Columbia University, 1255 Amsterdam Avenue, New York, NY 10027, USA}
\author{Melissa K. Ness}
\affil{Department of Astronomy, Columbia University, 550 West 120th Street, New York, NY, 10027, USA}
\affil{Center for Computational Astrophysics, Flatiron Institute, 162 Fifth Avenue, New York, NY, 10010, USA}

\begin{abstract}

Chemical abundances of stars in the Milky Way disk are empirical tracers of its enrichment history. However, they capture joint-information that is valuable to disentangle. In this work, we seek to quantify how individual abundances evolve across the present-day radius of the disk, at fixed supernovae contribution (\feh, \mgfe). We use 18,135 \apogee\ DR17 red clump stars and 7,943 \galah\ DR3 main sequence stars to compare the abundance distributions conditioned on (\feh, \mgfe) across $3-13$ kpc and $6.5-9.5$ kpc, respectively. In total we examine 15 elements: C, N, Al, K (light), O, Si, S, Ca,  ($\alpha$), Mn, Ni, Cr, Cu, (iron-peak) Ce, Ba (s-process) and Eu (r-process). We find that the conditional neutron capture and light elements most significantly trace variations in the disk's enrichment history, with absolute conditional radial gradients $\leq 0.03$ dex/kpc. The other elements studied have absolute conditional gradients $\lesssim 0.01$ dex/kpc. We uncover structured conditional abundance variations as a function of [Fe/H] for the low-$\alpha$, but not the high-$\alpha$ sequence. The average scatter between the mean conditional abundances at different radii is \sigint\ $\approx$ 0.02 dex (with Ce, Eu, Ba \sigint\ $>$ 0.05 dex). These results serve as a measure of the magnitude via which different elements trace Galactic radial enrichment history once fiducial supernovae correlations are accounted for. Furthermore, we uncover subtle systematic variations in all moments of the conditional abundance distributions that will presumably constrain chemical evolution models of the Galaxy. 

\end{abstract}

\section{Introduction}

The field of Galactic archaeology has advanced over the past decade as a result of many large spectroscopic surveys. The field has moved from having information for only hundreds of stars \citep[see e.g.][]{1993A&AS..102..603E,fuhrmann1998nearby,bensby2014exploring,2014Anders} to precise measurements of stellar paramters and individual abundances for more than $10^5$ stars. The third data release of the GALactic Archaeology with HERMES survey \citep[\galah;][]{Buder2021} provides stellar parameters and up to 30 abundance ratios for 588,571 stars, including information on 9 neutron capture elements. The Large sky Area Multi-Object fiber Spectroscopic Telescope \citep[LAMOST;][]{2012LAMOST,cui2012large}, accompanied with abundances by \cite{2019ApJS..245...34X}, provides 16 element abundances for 6 million stars, while the Gaia–European Southern Observatory (ESO) survey \citep[\Gaiaeso;][]{Gilmore2012} measures detailed abundances for 12 elements in about 10,000 field stars. The newly released DR17 of the Apache Point Observatory Galactic Evolution Experiment (\apogee) survey \citep{apogeeDR17,Majewski2017} provides insight into 20 abundance species for 657,135 stars, with significant improvements for the element cerium.

Despite the increase in spatial \citep[e.g.][]{2021Weinberg} and chemical abundance ranges \citep[e.g.][]{PriceJones2020} in surveys, and therefore a clearer picture of the current state of the Milky Way, there are still  fundamental questions that remain unanswered. For instance, the origin of the well known high- and low-$\alpha$ sequence bi-modality in the \feh--\alphafe\ plane has led to much debate, in which simulation work has proposed possible solutions \citep[e.g.][]{Clarke2019,2020Lian_alphaDichotomy,mackereth2018origin,2020_buckchemical,2017ApJ...837..183W,2021MNRAS.507.5882S}. Observationally, the high- and low-$\alpha$ sequences have different dynamical properties, even at fixed age \citep[e.g.][]{Gandhi2019,mackereth2019dynamical}. Many surveys have revealed that the relative fraction of high- and low-$\alpha$ stars varies depending on disk height and radius \citep[][]{2012bensby, hayden2015chemical, nidever2014tracing}. Primarily concentrated towards the center of the Milky Way, the high-$\alpha$ sequence is barely present in the outer disk, while the low-$\alpha$ sequence is present further outwards. This poses the question: did the few high-$\alpha$ stars of the outer disk migrate outwards from the bulk of the high-$\alpha$ stars in the inner disk, or has the sequence evolved across radii? While the relative effect of the radial mixing processes blurring and churning (radial migration) are still under debate, radial mixing is thought to be important in Milky Way evolution \citep{2020Feltzing,2016ApJ...818L...6L,2013A&A...558A...9M,2013A&A...553A.102D}. Given that we are only able to infer birth radius under some modeling assumptions \citep[e.g.][]{2018Minchev_rbirth,Frankel2018}, and never directly measure it, we are unable to directly test if these high-$\alpha$ stars migrated outwards.

Additional element abundances may be important to inform the link between the $\alpha$-bimodality and disk formation history. However, the tensions between yield tables, chemical evolution models, and data reduce our ability to understand the diversity of the sequences' enrichment histories \citep{blancato2019variations,2017A&A...605A..59R}. An important step forward is to map the empirical landscape of abundances to constrain theory and models. 

Theoretically, stars born at similar times and places should have similar chemical abundance trends, where the chemical homogeneity of star clusters is on the order of 1 pc \citep{BH2010,Armillotta2018}. Given that stellar orbits evolve over time and their dynamical properties change \citep[e.g.][]{Selwood2002, Roskar2008, 2009MNRAS.396..203S, 2010Michev, 2018Hayden_radialMigration, 2012MNRAS.426.2089R}, it is best to utilize the relatively unchanging chemical abundances of stars to determine stellar birth populations. This is the goal of chemical tagging \citep{2002freeman-BH}. While chemical tagging has promise  \citep{Hogg2016,2016martell,2022MNRAS.510.2407B}, there have been concerns over its need for large sample sizes ($>10^6$ stars) and high precision data \citep{Ting2015,2013Lindegren}. Recently, \cite{2022Ratcliffe} showed that for the size and precision of current Milky Way data sets, it is probable that groups of chemically similar stars  occupy different locations in birth time and place from one another. Thus, highlighting the utility of element abundances to differentiate overall birth environments, if not specific clusters.

In this paper, we perform a radial exploration of what we refer to as conditional abundances throughout the Milky Way disk, for a given contribution of supernovae type Ia (SNIa) and II (SNII), using [Fe/H] and [Mg/Fe] as the fiducial measures of these channels. That is, we wish to assess which elements capture information about the radially varying birth environment of the disk beyond bulk (\feh, \mgfe) measurements. While it is understood that the full chemical space collapses onto only a few dimensions \citep[][]{Ness2019,PJ2018,ting2012principal} and some work has used this to reduce the dimensions of the abundance space for analysis \citep[e.g.][]{casey2019data,garciaDias2019}, each element should in principle contribute uniquely to the chemical evolution history of the Galaxy \citep{Kobayashi2020}.

Recent work has also conditioned on \feh\ and \mgfe\ to understand the additional power other abundances have in resolving the Milky Way's evolution, overall. The two channels of (\feh, \mgfe) alone are able to well-predict [X/Fe] information \citep[e.g.][]{2022Ness}. However, seven abundances are needed in order to remove residual correlations, and each element may provide individual information linking to birth properties \citep{2021TingWeinberg}. Our work builds upon these studies. Here, however, we do a radial analysis of the additional information captured in abundances conditioned on supernovae contribution, in a model-free way. We seek to answer the following questions: (i) at fixed contribution of supernovae-generated elements (\feh, \mgfe), do abundance patterns evolve across present day radii, or are they unvarying and presumably therefore drawn from the same underlying chemical population; and (ii) which nucleosynthetic families vary the most across the disk conditioned on (\feh, \mgfe)? 

This work also extends upon the work of \cite{Weinberg2019} and \cite{2021Griffith}, who showed that abundance trends throughout the bulge and disk are independent of galactic location for the $\alpha$-sequences, and that a star's abundance pattern can be represented as a sum of contributions from SNIa and SNII.  \cite {2021Griffith_residuals} and \cite{2021Weinberg} extend the two-process model used in their previous work to \galah+\ DR3 and \apogee\ DR17 elements, and find that the correlations of the residuals reveal underlying structure such as common nucleosynthetic enrichment sources. In these recent works, the analyses were done on two groups by modeling the data for the $\alpha$-sequences separately. However, it has been suggested that empirically the Milky Way can be considered as a continuous evolution, and not in terms of only two chemical populations \citep{2020Ratcliffe,bovy2012milky}. Therefore, we separate our data into bins of \feh\ and \mgfe\ across the full \feh--\mgfe\ plane. We subsequently explore how the element abundance distributions change radially within these so-called chemical cells. We additionally separate the chemical cell populations as a function of age to explore how the individual abundance distributions of the disk changes for a given time period. 

This paper is organized as follows. In Section \ref{sec:data} we discuss the data sets used in this work. Our results are detailed in Section \ref{results}. The conditional abundance gradients across the Milky Way disk are detailed in Section \ref{sec:results I}. The conditional \xfe\ distributions for different radial regions in the disk are shown in Section \ref{sec:results II}. Finally, the conditional scatter and bias in abundances is demonstrated in Section \ref{sec:results III}.  Section \ref{sec:discussion} presents our discussions and key conclusions from this work. \\

\section{Data}\label{sec:data}

In order to capture element abundances from a variety of nucleosynthetic families throughout the Milky Way disk, we use data sets from two surveys for this work --- \apogee\ and \galah. 

\subsection{\apogee\ DR17}\label{sec:data_lucy} 

\begin{figure*}
     \centering
     \includegraphics[width=\textwidth]{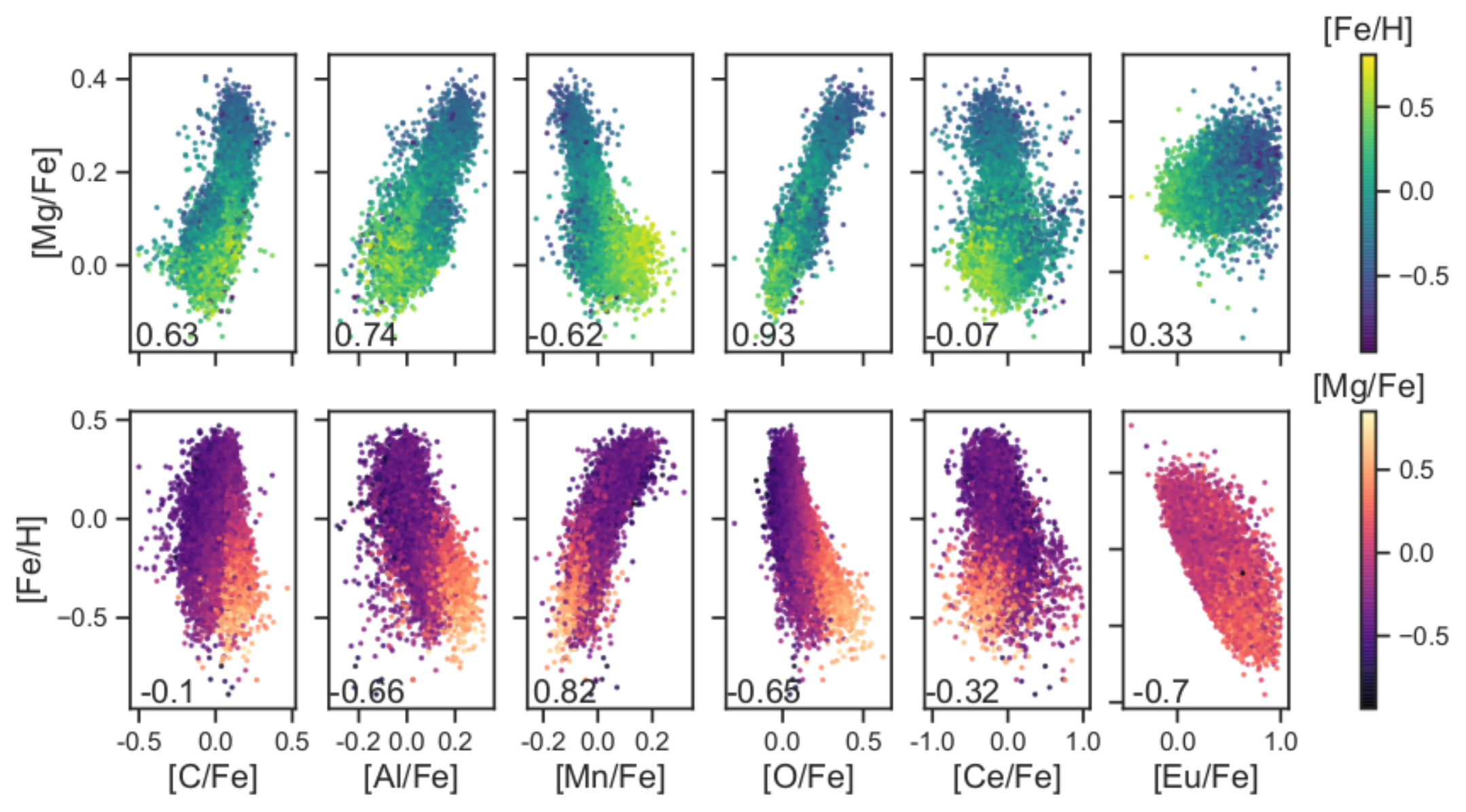}\\
\caption{\xfe--\mgfe--\feh\ relationships for 18,135 \apogee\ stars and 7,943 \galah\ stars. \cfe, [Al/Fe], [Mn/Fe], [O/Fe], and [Ce/Fe] are measured by \apogee, and \eufe\ is measured by \galah. The correlation between \xfe\ and \mgfe\ (top row) and \feh\ (bottom row) is given in the lower left hand corner of each panel. There are inter- and intra- familial correlations, showing that (\feh, \mgfe) capture a wide variety of abundance information.}
\label{fig:correlations}
\end{figure*}

In our analysis, we use 14 abundances (\feh, \xfe\ for Mg, C, N, Al, K, O, Si, S, Ca, Ce, Mn, Ni, Cr) provided by \apogee\ DR17 \citep{apogeeDR17, Majewski2017}, the final release from the fourth phase of the Sloan Digital Sky Surveys \citep[SDSS-IV;][]{blanton2017sloan}, processed by the \apogee\ Stellar Parameter and Chemical Abundance Pipeline \citep[\aspcap;][]{GP2016}. We partnered these abundances with the age catalogue of \cite{yuxi2021universal}, where ages were derived with an uncertainty of $\sim$1.5 Gyr from their spectra using \thecannon\ \citep{Ness2015}. Galactic radius (\rgal) was derived from distances by \cite{2018AJ....156...58B} using \Gaia\ DR2 \citep{gaiaDR2} parallaxes, and assuming the sun is at 8.2 kpc.  \rgal\ has a median uncertainty of $\sim$0.03 kpc.

\begin{figure*}
     \centering
     \includegraphics[width=.95\textwidth]{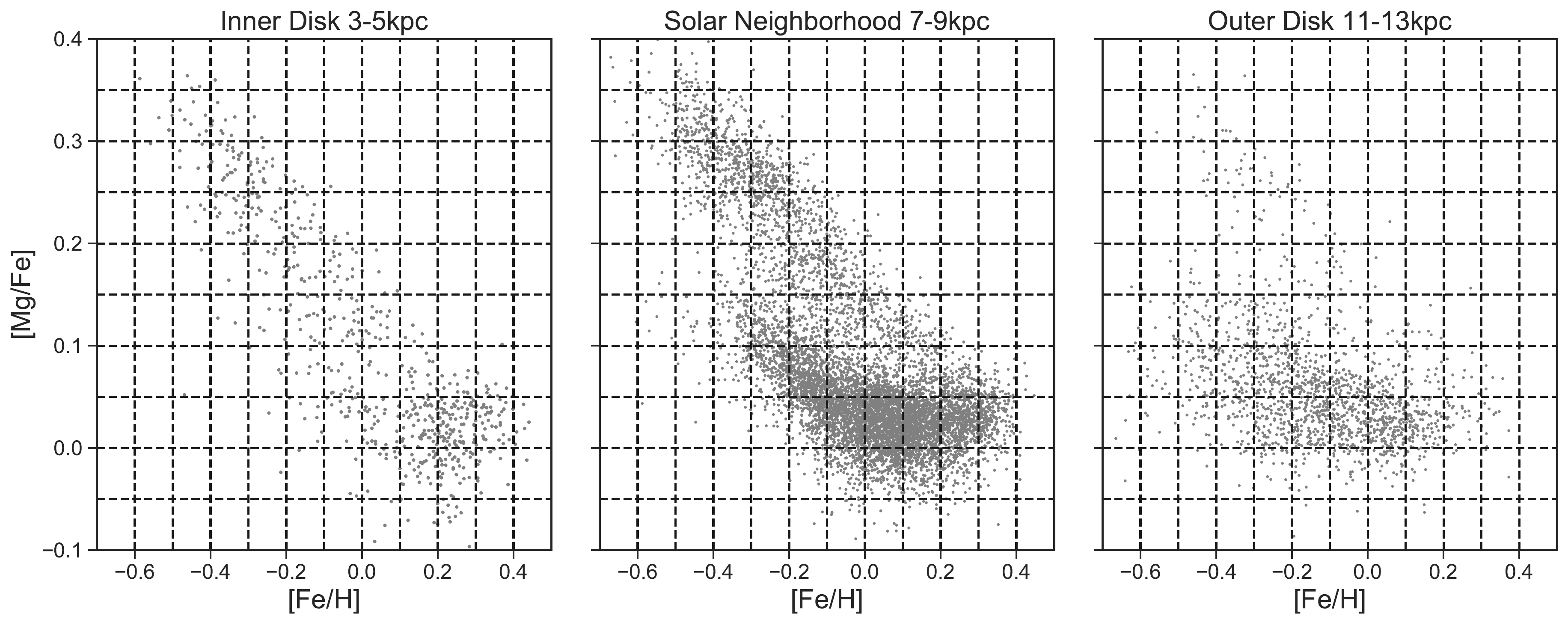}\\
     \includegraphics[width=.6\textwidth]{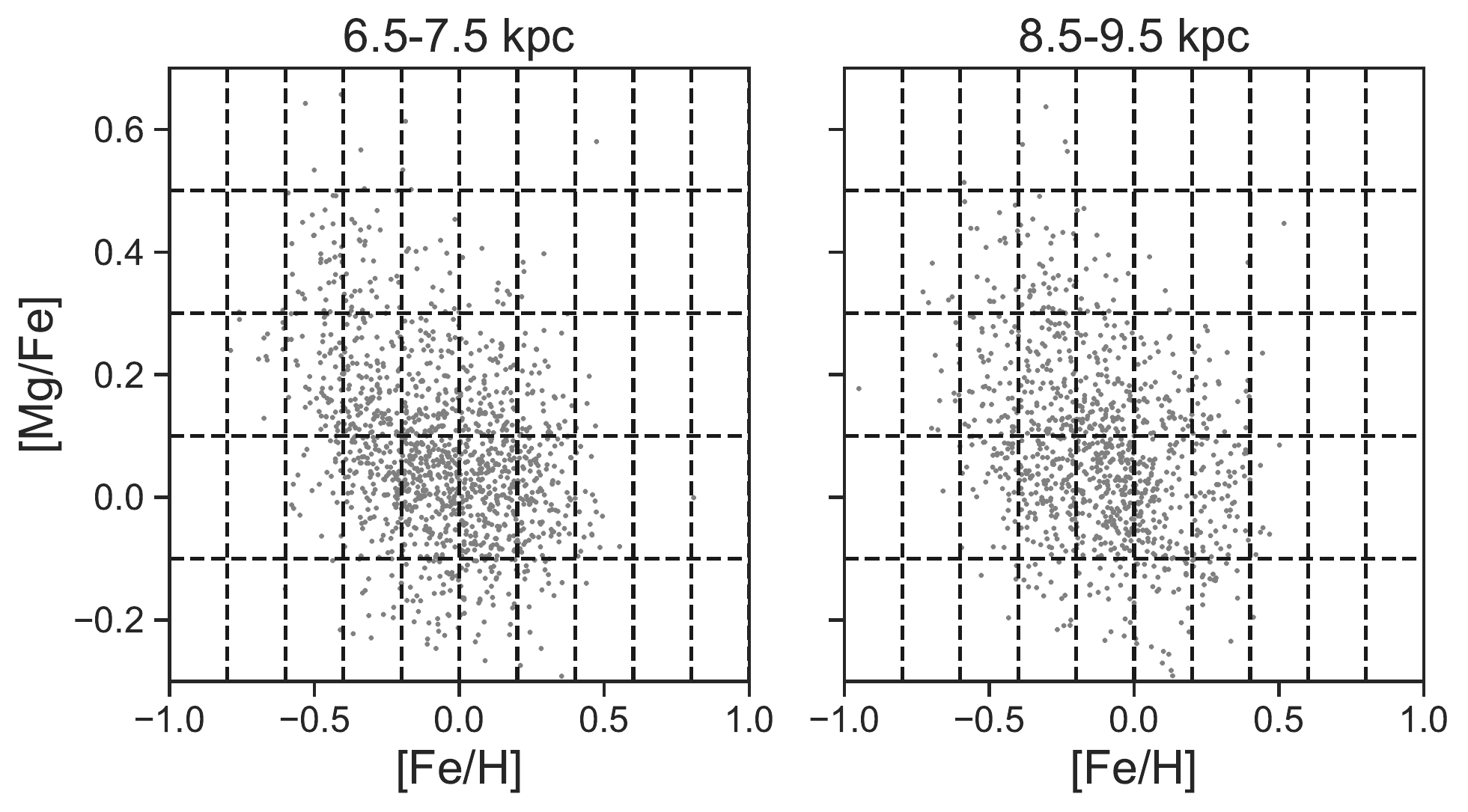}\\
\caption{The \feh--\mgfe\ plane of our stellar sample. (\textbf{Top}): 9,736 red clump \apogee\ DR17 stars and (\textbf{Bottom}): 2,418 main sequence \galah\ stars), overlaid by the chemical cells used in analysis, separated into different radial regions, indicated as the top of each panel. The \apogee\ inner disk has 630 stars, the \apogee\ solar neighborhood has 7,501 stars, and the \apogee\ outer disk has 1,605 stars. The \galah\ inner and outer solar neighborhoods have 1,349 and 1,069 \galah\ stars respectively.}
\label{fig:data}
\end{figure*}

In order to avoid systematic abundance trends with evolutionary state induced in abundances by both approximate analyses \citep[e.g.][]{Jofre2016}, as well as  astrophysical processes like diffusion and dredge-up \citep{Liu2019,2017ApJ...840...99D}, we focus our analysis on the \apogee\ red clump. \cite{yuxi2021universal} determines red clump candidates from their spectra, using data-driven modeling of the correlation between flux variability and evolutionary state \citep{Hawkins2018}. They report a  contamination rate of 2.7\%. 

To remove outliers with highly anomalous (and likely not astrophysical) abundance measurements from our analysis, we only consider stars with abundances $-1 \leq \xfe \leq 1$. We also ensure a high quality sample by keeping stars with unflagged abundances (\code{X\_FE\_FLAG = 0} and \code{FE\_H\_FLAG = 0}) and a signal to noise ratio $>$ 50. This leaves us with a sample size of 18,135 stars from the \apogee\ survey for analysis of 12 element abundances from 5 nucleosynthetic channels --- 3 $\alpha$-elements (S, Si, Ca, O), 3 iron peak (Ni, Mn, Cr), 3 light (C, N), 2 light-odd Z (K, Al), and 1 s-process (Ce). The mean measurement uncertainty across these elements is $\sim$0.03 dex, with the largest uncertainty being in [Ce/Fe] (0.08 dex).

\subsection{\galah\ DR3 }\label{sec:data_galah}

We additionally look at regions closer to the solar neighborhood for \galah\ DR3 \citep{Buder2021} disk ($|z|\leq1$ kpc) stars. Using these stars we similarly explore the conditional abundance trends of Ba and Eu, along with the elements C, K, Al, Ca, Si, Mn, Ni, and Cu. Stellar ages are determined using Bayesian Stellar Parameter Estimation code \citep[BSTEP;][]{2018bstep}, by making use of stellar isochrones and a flat prior on age and metallicity. The mean age uncertainty is 2 Gyr. Galactocentric radii are provided in the dynamics value added catalog, with a median uncertainty of $\sim$0.01 kpc.

For the \galah\ data, we choose to focus on a narrow range in \logg\ and temperature, which we define as stars with $3.5 \leq \logg \leq 4$ and $5,000$ K $\leq \teff \leq 5,500$ K, to mitigate any effects due to stars in different evolutionary states. Similar as done with the \apogee\ data, we remove stars that have absolute abundance measurements greater than 1. Our \galah\ sample consists of 7,943 stars with 10 abundances across six nucleosynthetic families --- 2 $\alpha-$elements (Ca, Si), 3 iron peak (Mn, Ni, Cu), 1 light (C), 2 light-odd Z (K, Al), 1 r-process (Eu), and 1 s-process (Ba). The average measurement uncertainty is $\sim$0.11 dex.

\subsection{Chemical cells}
Figure \ref{fig:correlations} shows the \xfe--\feh--\mgfe\ relationships between six abundances used in this work (one from each nucleosynthetic family). There are strong correlations between \feh\ and [Al/Fe], [Mn/Fe], [O/Fe] and [Eu/Fe], while [C/Fe], [Al/Fe], [Mn/Fe], and [O/Fe] have strong correlations with \mgfe. While these (\feh, \mgfe) correlations capture the bulk of \xfe\ information, we wish to examine the abundance information not captured in the correlations. 

In this work, we examine the \xfe\ abundance patterns of individual elements in chemical cells. In particular, we investigate the mean radial trends of element distributions from \rgal\ = $3-13$ kpc conditioned on supernovae type Ia and II contribution (represented by the most precisely measured elements \feh\ and \mgfe). Using \apogee\ data, we compare in detail the conditional element distributions for the inner disk (\rgal\ = $3-5$ kpc), solar neighborhood (\rgal\ = $7-9$ kpc), and outer disk (\rgal\ = $11-13$ kpc) of 9,736 stars observed within those regions. We define bin sizes of 0.1 dex in \feh\ and 0.05 dex in \mgfe, well above the mean \apogee\ measurement uncertainty of 0.009 dex and 0.012 dex for \feh\ and \mgfe, respectively. The top row of Figure \ref{fig:data} shows our \apogee\ selection in the three radial regions, with the Cartesian grid defining the chemical cells overlaid. In Section \ref{sec:results III} we additionally condition on stellar age with bins $0-3$ Gyr, $3-6$ Gyr, $6-9$ Gyr, and $9-12$ Gyr.

Since \galah\ surveys closer to the solar neighborhood, for our analysis involving this data we focus on the 2,418 stars in the inner solar neighborhood (\rgal\ = $6.5-7.5$ kpc) and outer solar neighborhood (\rgal\ = $8.5-9.5$ kpc). The average \galah\ measurement uncertainty is 0.1 dex in \feh\ and 0.13 dex in \mgfe. As this is larger than the \apogee\ sample, we define correspondingly larger bin widths for the \galah\ sample, of 0.2 dex in both \feh\ and \mgfe. The bottom row of Figure \ref{fig:data} shows the \galah\ sample and chemical cell divisions.

\section{Results}\label{results}

Using the chemical cells across the [Fe/H]--[Mg/Fe] plane (as demonstrated in Figure \ref{fig:data}, for different radial bins), we first examine the conditional abundance gradients for the 12 elements in \apogee\ across \rgal\ = $5-13$ kpc and the 10 elements in \galah\ across \rgal\ = $7-9$ kpc (Section \ref{sec:results I}). This demonstrates the relative behavior and magnitude of gradients both within and between elements, for different (\feh, \mgfe) cells. Then, we look beyond the mean measure of the conditional abundance gradients, to the conditional abundance distributions at different radii, using our radial bins described in section \ref{sec:data}. In particular, we inspect the variation of the four statistical moments of the conditional distributions; the mean, standard deviation, skew, and kurtosis (Section \ref{sec:results II}). We end our analysis by quantifying the overall mean conditional abundance bias and scatter across Galactic radii, for each element, for all \feh--\mgfe\ chemical cells together (Section \ref{sec:results III}). This enables us to make direct comparisons to existing literature and to demonstrate how the higher resolution information seen in the gradients is captured on average across the disk. 

\subsection{I: Conditional abundance gradients across Galactic radii} 
\label{sec:results I}
 
\begin{figure}
      \includegraphics[width=0.5\textwidth]{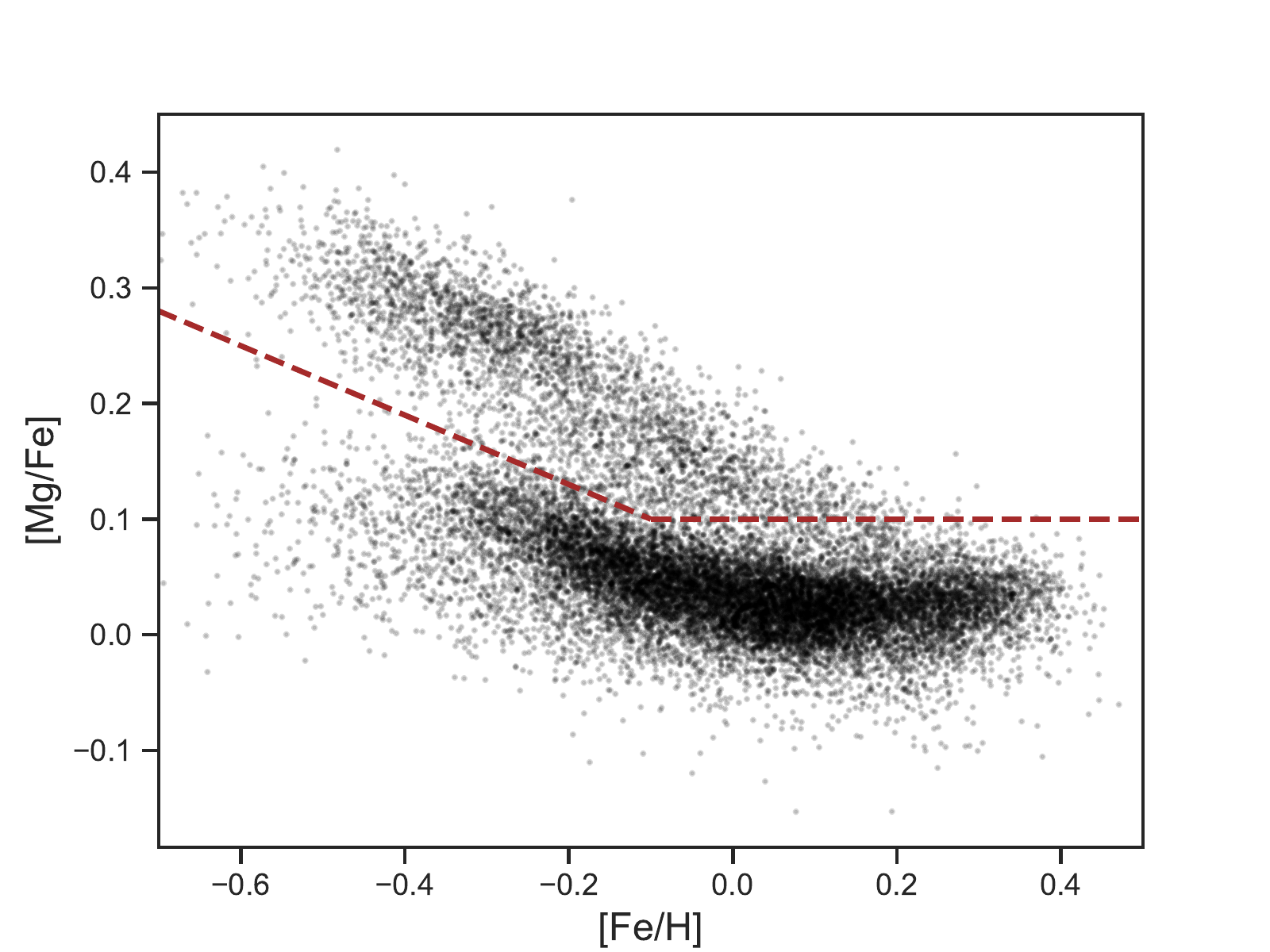}
\caption{18,135 \apogee\ DR17 stars in the \feh--\mgfe\ abundance plane. The dashed brown line represents our simple split into the high- and low-$\alpha$ sequence for visual plotting purposes in Section \ref{sec:results I}.}
\label{fig:alphaSeqs}
\end{figure}

\begin{figure*}
     \centering
      \includegraphics[width=1\textwidth]{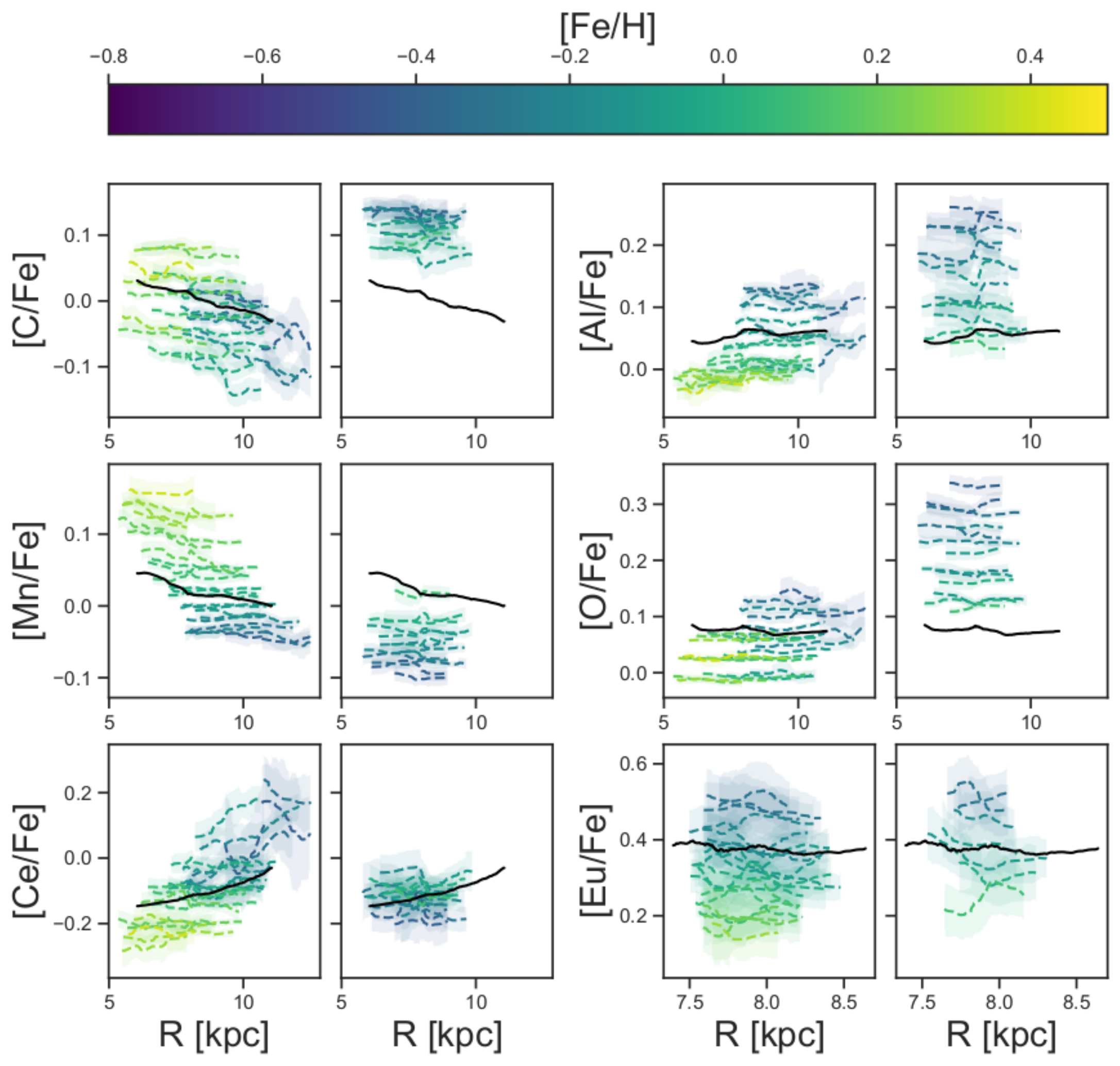}
\caption{[X/Fe] running mean (dashed line) as a function of \rgal\ for each \feh--\mgfe\ chemical cell with at least 100 stars, colored by the cell's mean [Fe/H]. The ribbon around the running mean represents the 95\% confidence on the mean. The running means are split into 2 figures for each element, where bins in the low-$\alpha$ sequence are on the left, and bins in the high-$\alpha$ sequence are on the right (separation of high- and low-$\alpha$ sequences is shown in Figure \ref{fig:alphaSeqs}). The black solid line shows the overall running mean of \rgal--\xfe. Here, \cfe, [Al/Fe], [Mn/Fe], [O/Fe], and [Ce/Fe] are measured by \apogee, and \eufe\ is measured by \galah. For most abundances, the overall running mean does not capture the subtleties of the gradient as seen for a given value of (\feh, \mgfe).}
\label{fig:runningMean}
\end{figure*}

We examine the running means of elements across a radial range of $\Delta$\rgal\ = 8 kpc for each (\feh, \mgfe) cell. We set the radial bin size to calculate the running mean over 20\% of the stars in each chemical cell at each step. We then smooth the conditional gradients across radius using a Gaussian filter with a standard deviation of $\sigma$ = 4 and a filter size of 9. Figure \ref{fig:runningMean} shows the running means for 6 elements, selected to showcase the different nucleosynthetic families: \cfe\ (light), [Al/Fe] (light odd Z), [Mn/Fe] (iron peak), [O/Fe] ($\alpha$), \cefe\ (s-process), and \eufe\ (r-process). The shaded regions represent the 95\% confidence interval on the mean, and are colored according to the cell's \feh\ value. We also show for reference the mean of the full set of stars (not binned in [Fe/H]--[Mg/Fe]) with a black line. The conditional abundance gradients are, in general, flatter than the original gradients, and show variability as a function of chemical cell. 

\begin{figure*}
     \centering
     \includegraphics[width=1\textwidth]{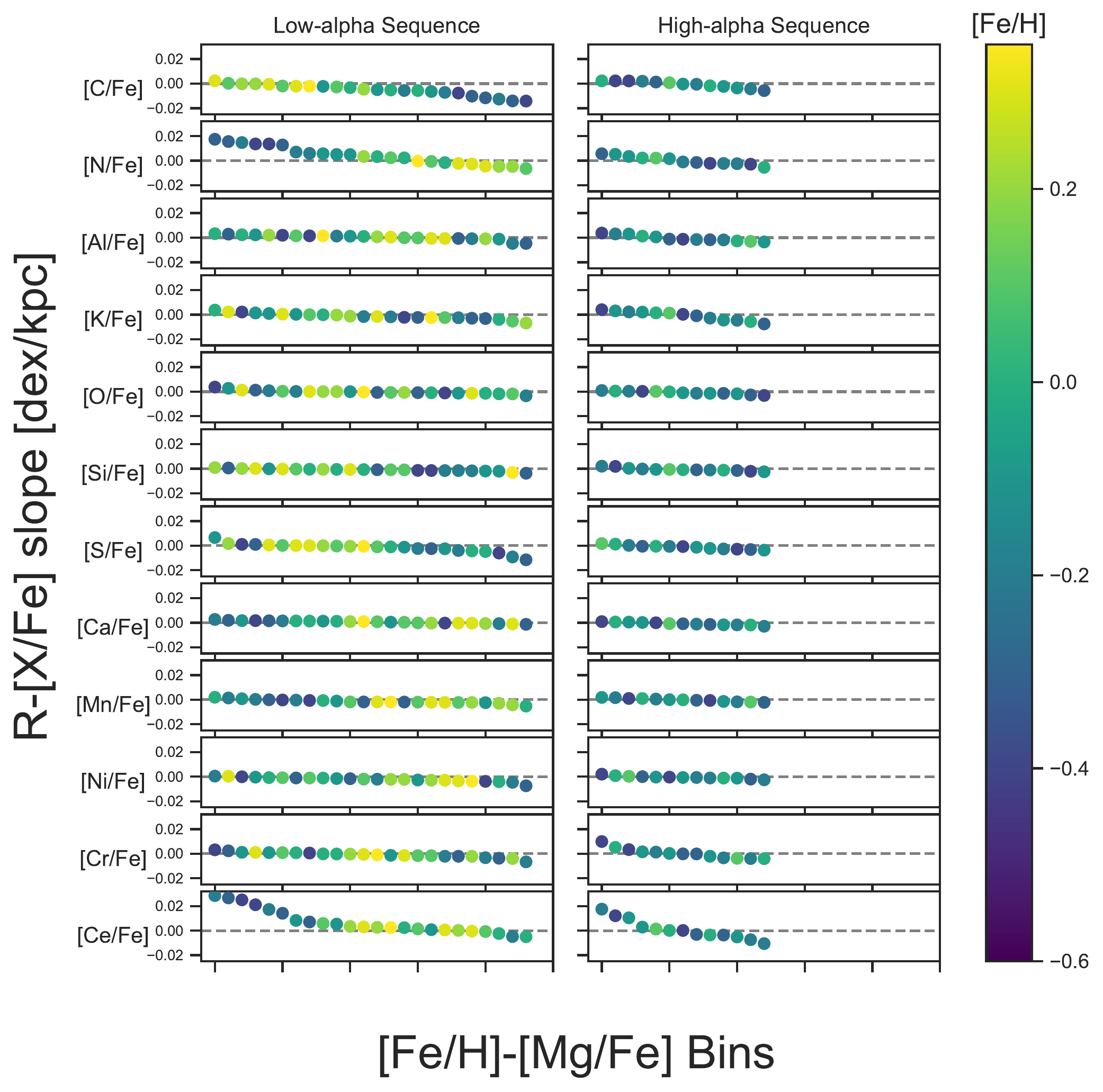}
\caption{[X/Fe]--\rgal\ slopes for different \feh--\mgfe\ chemical cells with at least 100 stars for abundances measured in \apogee\ separated by the cell's classification in the high- or low-$\alpha$ sequence. These are ordered in both panels from the most positive to most negative slopes. Most cells in the high-$\alpha$ sequence have a fairly flat slope, with most of the \xfe\ variation in \rgal\ coming from the low-$\alpha$ sequence, at the lower [Fe/H] values of this distribution. The elements \cefe, [N/Fe], and [C/Fe] have the largest gradients with \rgal\ seen in a subset of chemical cells.}
\label{fig:apogee_slopes}
\end{figure*}

Specifically, and starting with \cfe\ (top left panels of Figure \ref{fig:runningMean}), we can see from the running mean of the full sample that \cfe\ has a negative, fairly steep gradient in the Milky Way disk \citep[in agreement with literature;][]{Eilers2022}. However, for a given value of \feh\ and \mgfe, the amplitude of the conditional gradient changes as a function of \feh. For smaller values of \feh, the \rgal--\cfe\ conditional gradients are steeper than gradients of the higher valued \feh\ chemical cells, where most of the slopes are nearly flat. In fact, for some of the chemical cells with the highest value of \feh\ (0.4 dex) we can see a very slight positive correlation between \cfe\ and \rgal.

Similar to the \rgal--\cfe\ gradients, the \rgal--[Mn/Fe] (middle left panels), --[O/Fe] (middle right panels), and low-$\alpha$ --[Al/Fe] (top right panels),  conditional abundance gradients for each chemical cell have small uncertainties about the running mean. Even though the overall gradient in these abundances is slightly positive ([Al/Fe]), negative ([Mn/Fe), or flat ([O/Fe]), the conditional gradients in chemical cells for these elements overall have smaller slopes in \rgal. The one exception is the \feh--\mgfe\ chemical cells in the high-$\alpha$ sequence, where [Al/Fe] has a jump in abundance near the solar neighborhood. 

While the gradient trends of the light, $\alpha$-, and iron peak elements are primarily linear for each \feh--\mgfe\ cell, that is not true for other abundances. The chemical cells of \cefe\ and \eufe\ in the bottom row of Figure \ref{fig:runningMean} show that the relationship between \xfe\ and \rgal\ can be complex. While the overall relationship between \cefe\ and \rgal\ is a simple positive correlation with a consistent slope, the gradients of \cefe\ after conditioning on \mgfe\ and \feh\ are intricate. The high-\feh\ chemical cells have a linear relationship between \cefe\ and \rgal, while lower-\feh\ cells ($<0.2$ dex) have a steeper, and somewhat quadratic relationship. The running means of \eufe\ versus \rgal\ also show this quadratic, and sometimes cubic relationship between \eufe\ and \rgal\ for \feh--\mgfe\ populations.

In order to quantify the \rgal--\xfe\ conditional relationship, we fit a linear model estimating \xfe\ from \rgal\ in each chemical cell. Figure \ref{fig:apogee_slopes} shows the slopes from the regression for the elements measured in \apogee. In this figure we sort the slopes for each element from largest to smallest, to showcase the range. Similar to Figure \ref{fig:runningMean}, we split the chemical cells into high- and low-$\alpha$ groups for each element and color by \feh\ to understand the influence of \mgfe\ and \feh\ on the gradients. 
Looking at Figure \ref{fig:apogee_slopes}, we see that almost all \feh--\mgfe\ chemical cells classified in the high-$\alpha$ sequence have a near-zero slope, and that the informative (non-zero gradient conditional abundances) chemical cells are in the low-$\alpha$ sequence. The biggest exceptions to this are \cefe\ and [K/Fe], where high-$\alpha$ sequence bins have negative slopes, and [Cr/Fe], where some of the high-$\alpha$ chemical cells have steep positive slopes.

We also note that the conditional gradients of the light elements ([C/Fe] and [N/Fe]) have relatively ordered trends with \feh\ in the low-$\alpha$, but not high-$\alpha$ sequence. In the low-$\alpha$ sequence, cells with higher \feh\ (0.2 to 0.35 dex) have flatter abundance gradients than cells with lower \feh. The slopes of \cefe\ in the low-$\alpha$ sequence also differ as a function of \feh, where cells with high \feh\ ($>0.2$ dex) have near zero conditional gradients, whereas chemical cells with \feh\ values $<0.02$ dex have much steeper conditional gradients. We note that iron-peak elements Ni and Cr do not have strong trends with \feh\ despite being produced during SNIa. This is perhaps explained by their weak correlation with \feh\ \citep[e.g.][]{2020Ratcliffe}. However, other abundances with strong correlations with \feh\ (e.g. [Si/Fe]) also do not show any obvious trends with \feh. Overall, the $\alpha$-elements have the weakest slopes out of the nucleosynthetic families shown, suggesting they are not very informative about birth location beyond bulk \feh--\mgfe\ grouping. 

\subsubsection{Conditional abundance gradients in different age ranges}
\label{sec:slopes_age}

\begin{figure*}
     \centering
     \includegraphics[width=1\textwidth]{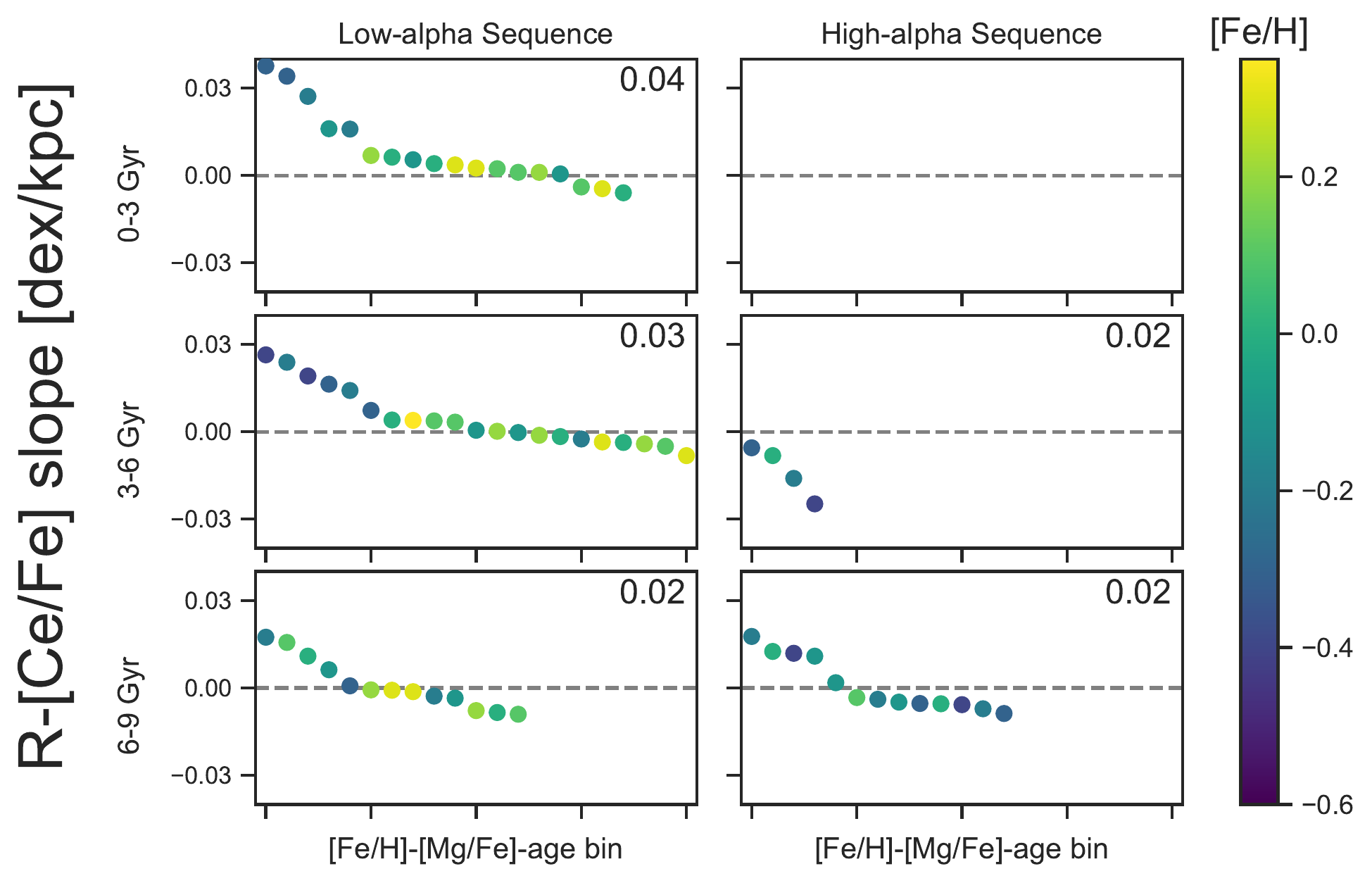}
\caption{\rgal--\cefe\ slopes for \feh--\mgfe\ chemical cells with at least 50 stars separated by low- or high-$\alpha$ sequence classification for 3 different stellar age groups. The largest absolute slope is given in the top right corner of each panel. The steepest gradients come from younger ($<6$ Gyr) stellar bins. Across all elements (except one chemical cell for [K/Fe]), the absolute slope of the chemical-age cells decreases as stellar age increases.}
\label{fig:cefe-slope}
\end{figure*}

In the previous section, we showed that in general, \xfe\ gradients flatten when conditioned on \feh\ and \mgfe, with the lowest metallicity (\feh\ $\leq -0.5$ dex) chemical cells showing the most variability as a function of radius. These low metallicity chemical cells also correlate with the largest age gradients, therefore we now test to see if the age dispersion is driving the results.

Since the Milky Way disk's abundance gradients change over time \citep{2001ApJ...554.1044C}, and we know that radial migration is strong and a function of age \citep[e.g.][]{Frankel2018}, we might expect the conditional gradients to differ as a function of age. We therefore separate the chemical cells into different age bins ($0-3$ Gyr, $3-6$ Gyr, and $6-9$ Gyr) to investigate how trends may differ for different age groups. In this section we examine how the slope of the \xfe\ gradient changes when considering stellar age. We focus on the \rgal--\cefe--age relationship  as \cefe\ shows the largest overall range in conditional abundance gradient in Figure \ref{fig:apogee_slopes}.

Figure \ref{fig:cefe-slope} shows the conditional  \cefe\ slope for each \feh--\mgfe\ cell for the three age bins, and separated by high- and low-$\alpha$ sequence classification. In Figure \ref{fig:apogee_slopes}, and discussed in the previous section, some of the high-$\alpha$ chemical cells have a negative \rgal--\cefe\ slope, and some have a positive relationship. When additionally conditioning on age, we still see the negative gradients in the high-$\alpha$ sequence ($3-6$ Gyr). However, we also see that some bins with stars $6-9$ Gyr in age have a positive correlation between \cefe\ and \rgal. 

In the low-$\alpha$ sequence, we can see that the conditional \rgal--\cefe\ gradients flatten for older stars. The youngest age bin ($0-3$ Gyr) has the strongest positive correlation between \cefe\ and \rgal\ conditioned on (\feh, \mgfe), and contributes the most to the strong positive overall \cefe\ abundance gradient. Here, we only show the result for [Ce/Fe], however, the absolute maximum slope of the conditional gradients decreases as stellar age increases for nearly all conditional abundances. The conditional gradients for the $\alpha$- and iron peak elements show slopes of $\leq0.01$ dex within all age bins. 

\subsection{II: Conditional abundance distributions across Galactic radii} 
\label{sec:results II}

The mean conditional abundance trends across radius measures the overall additional resolving power in each element in capturing the radially-dependent changing star formation history of the disk.  In this section we go beyond the mean of the distribution of each element, to investigate how the element distributions themselves vary across radius for a given value of \mgfe\ and \feh. 

\begin{figure*}
     \centering
     \includegraphics[width=.9\textwidth]{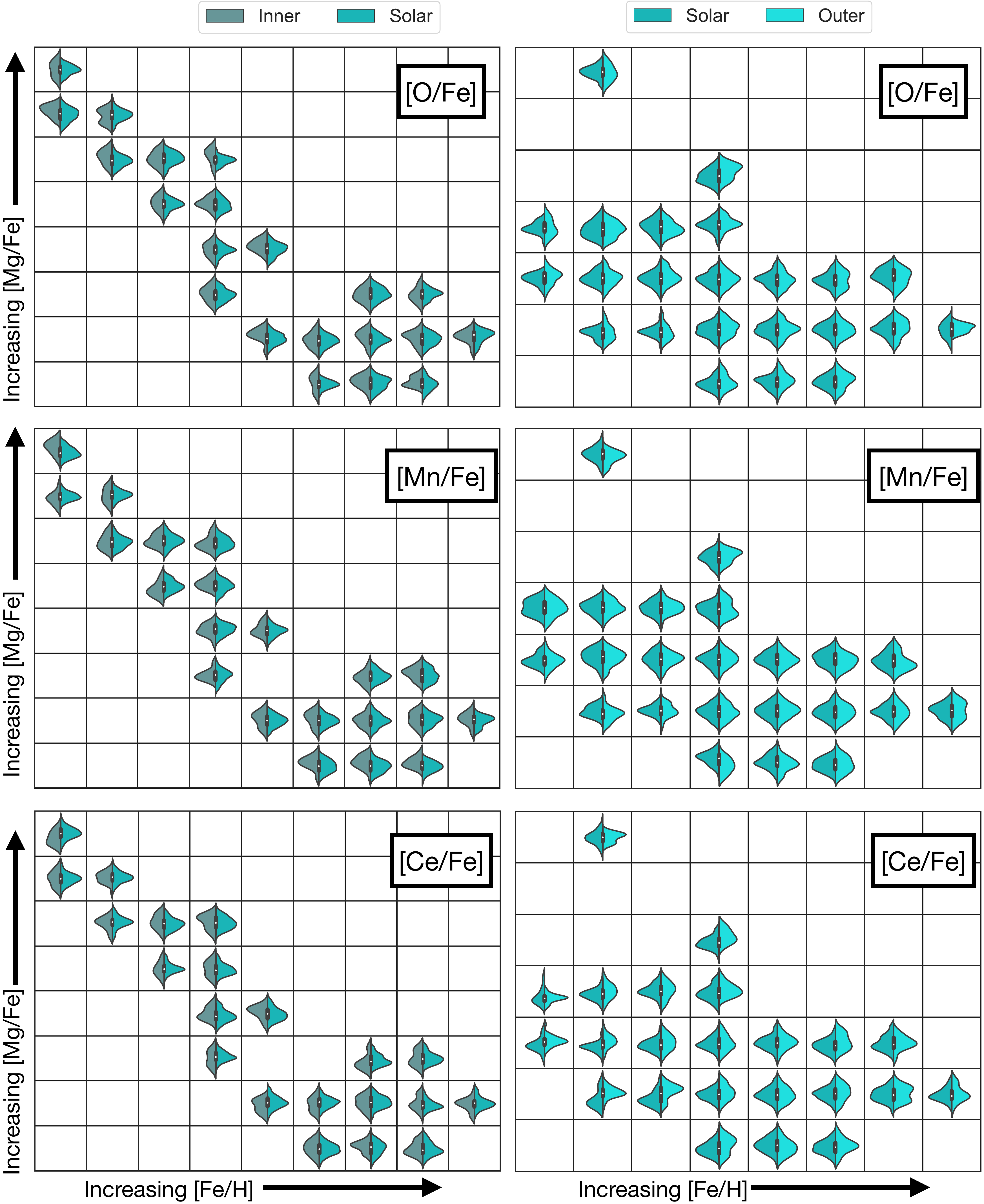}\\
     \caption{Split violin plots showing the \xfe\ density distribution for different \feh--\mgfe\ chemical cells in the inner disk vs solar neighborhood (left column) and solar neighborhood vs outer disk (right column). Here we show a representative element from the $\alpha$ (O), iron peak (Mn), and neutron capture (Ce) family. Outliers have been removed to more easily examine the key details. We can see that the distribution (such as modality) changes not only from cell to cell, but also across Galactic radii.}
\label{fig:violins2}
\end{figure*}

\begin{figure*}
     \centering
     \includegraphics[width=\textwidth]{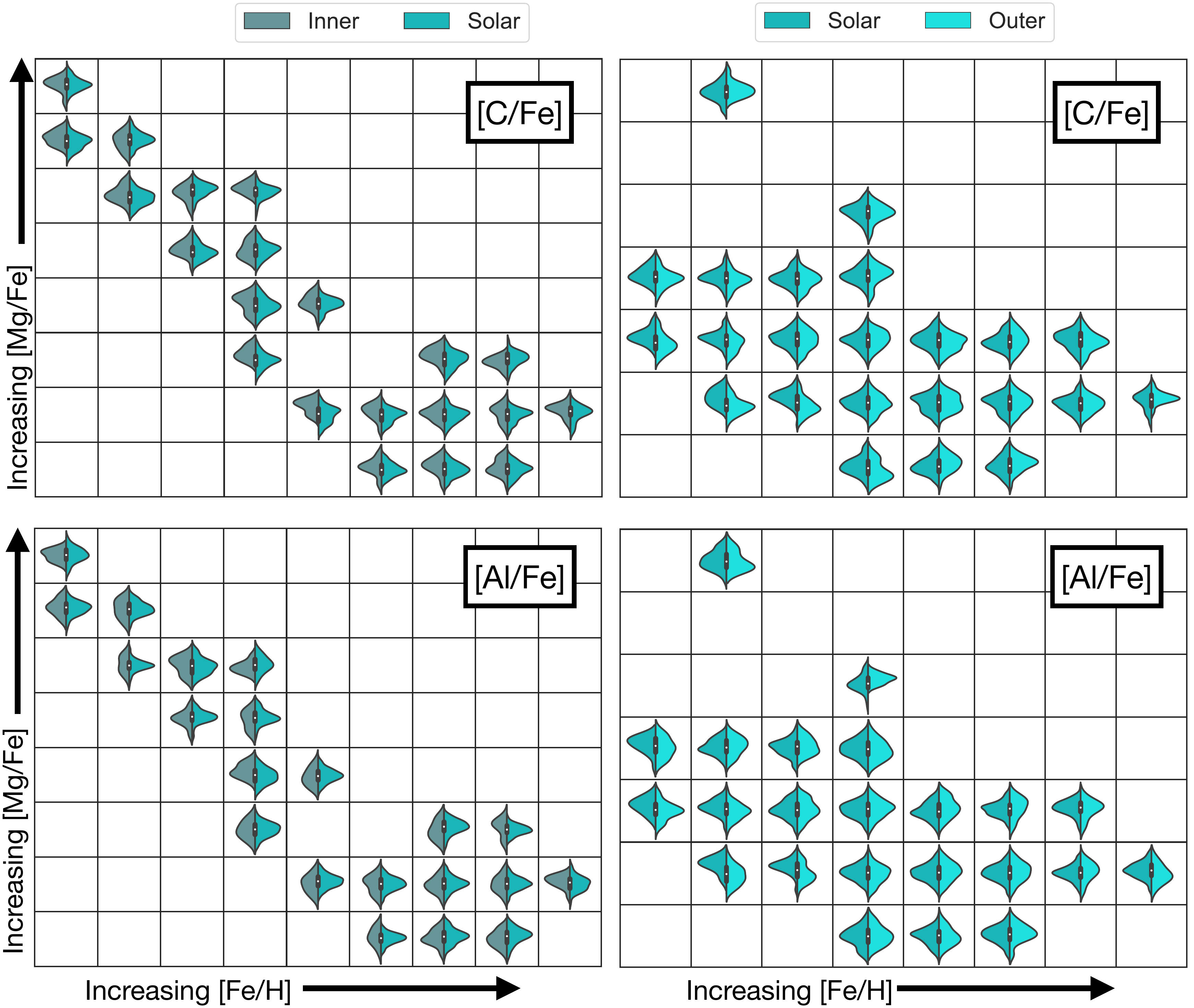}\\
     \caption{Similar to Figure \ref{fig:violins2}, here we show the abundance distributions for the light (C) and light odd Z (Al) nucleosynthetic family.}
\label{fig:violins1}
\end{figure*}

Figures \ref{fig:violins2} and \ref{fig:violins1} present split violin plots demonstrating the \xfe\ abundance distributions for the inner (\rgal\ = $3-5$ kpc), solar (\rgal\ = $7-9$ kpc), and outer (\rgal\ = $11-13$) regions of the Milky Way for \feh--\mgfe\ chemical cells that contain at least 10 stars. Similar to Section \ref{sec:results I}, we show a representative element from the light (C), light odd Z (Al), iron peak (Mn), $\alpha$ (O), and neutron capture (Ce) families. Since the r-process (Eu) is only available for the solar neighborhood, it is given in the appendix.

\begin{figure*}
     \centering
     \includegraphics[width=.92\textwidth]{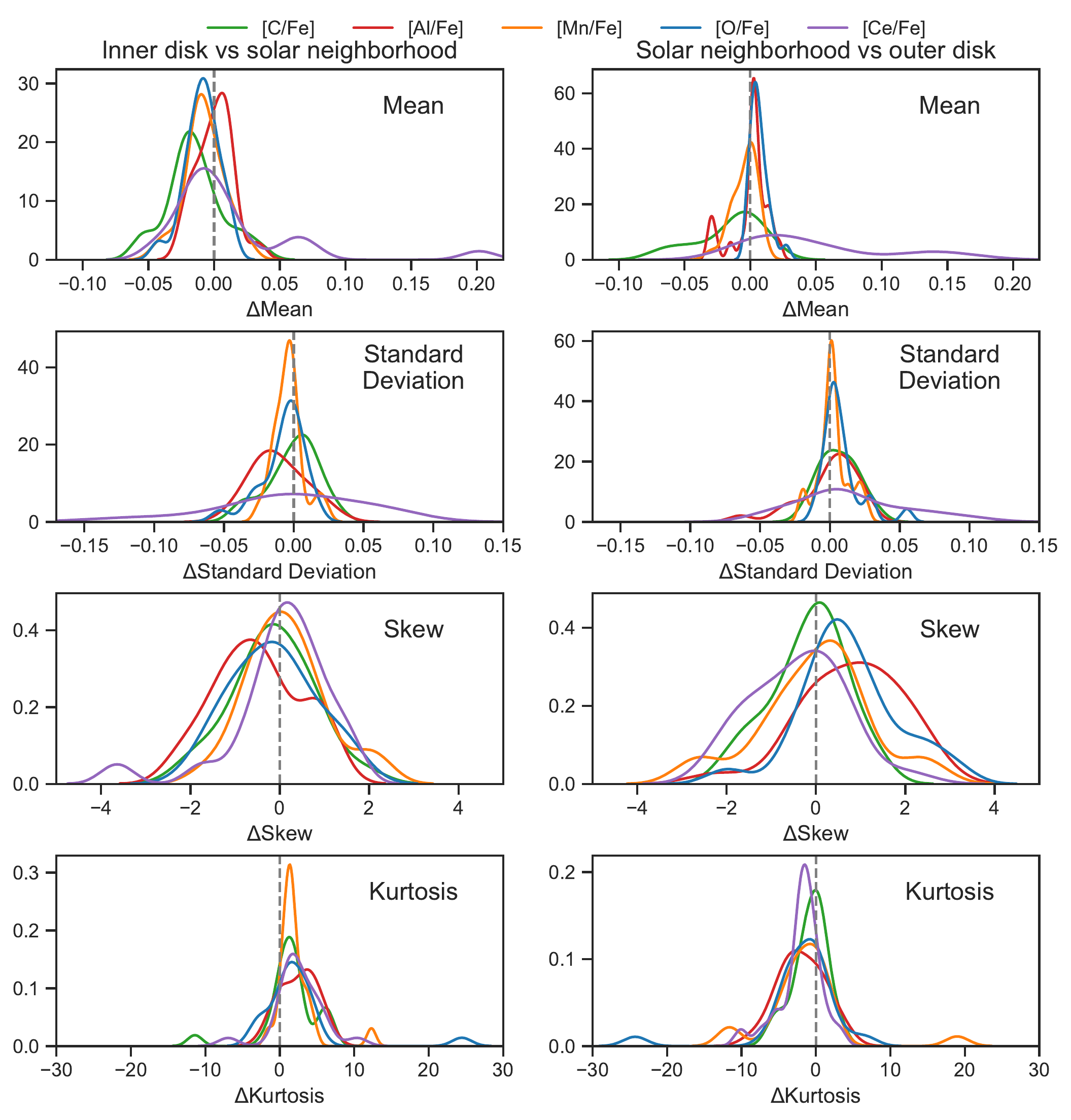}\\
     \caption{Density distributions of the differences in the statistical moments of the conditional abundances between radial bins. Each row shows a different moment, as labelled, where the value is calculated for each (\feh, \mgfe) cell, and combined to make each distribution. The left-hand panels compare the inner disk and solar neighborhood and the right-hand panels compare the solar neighborhood and outer disk,  for the five elements shown in Figures \ref{fig:violins2} and \ref{fig:violins1}. The dashed line at 0 indicates the reference for no difference in moments between the two radial regions. A positive net difference means the outer radial region has a larger value than the inner radial region and a negative net difference means the inner radial region has a larger value than the outer radial region. The elements [Al/Fe], [Mn/Fe], and [O/Fe] show little variation in distribution across Galactic radii, whereas \cfe\ has larger radial differences. \cefe\ shows the largest variation overall here, especially looking outwards in the disk.}
\label{fig:Moments}
\end{figure*}

In Figure \ref{fig:runningMean}, we saw that \cefe\ has a positive gradient in \rgal\ --- especially when looking outwards in the disk --- while the other abundance gradients became nearly flat when conditioning on \feh\ and \mgfe. This gradient in \cefe\ can be seen in some of the violin plots in the bottom row of Figure \ref{fig:violins2}---in particular the lower \feh\ cells of the solar vs outer neighborhood---where the distribution of the solar neighborhood is shifted downwards compared to the distribution of the outer disk. We can also see \cfe's decreasing trend in Galactic radii, especially for some of the lower \feh, lower \mgfe\ chemical cells. The distribution of the difference in \xfe\ means for each chemical cell is given as the top panels of Figure \ref{fig:Moments}, with the left panel showing the comparison between the inner disk and solar neighborhood, and the right panel showing the difference between the solar neighborhood and outer disk. This figure confirms that the mean increase in \cefe\ is concentrated towards the outer disk and that \cfe\ has a consistent negative gradient across most \feh--\mgfe\ cells. This figure also gives insight into how the gradients are distributed --- we can see that the distribution of the difference in \cfe\ abundance between the different radial regions is non-Gaussian and bi-modal. The details in these distributions are a potential measure of the detailed star formation and migration history. A similar bi-modality is definitively shown in [Al/Fe] and hinted at in [Mn/Fe] for the outer parts of the disk. The majority of the chemical cells show [O/Fe] and [Mn/Fe] have minor (0.01 dex) differences in mean abundance across radii.

Figures \ref{fig:violins2} and \ref{fig:violins1} also show that the scatter in \xfe\ is dependent on Galactic radii. For instance the distribution of [Al/Fe] in the inner disk of cell (3,2) has a larger scatter than in the solar neighborhood. To quantify these difference, the second row of Figure \ref{fig:Moments} shows the intrinsic difference of the standard deviations across radial regions for each of the \xfe\ distributions conditioned on \feh\ and \mgfe. To account for increased broadening in the \xfe\ distributions due to larger measurement uncertainty at further distances, we subtract the mean measurement uncertainty in the radial region in quadrature from the standard deviation. Similar to the difference in means, the standard deviation in \cefe\ varies widely across Galactic radii. The other abundances show the solar neighborhood has a slightly smaller scatter compared to the inner and outer regions of the Galaxy, however the scatter between different radial regions is more consistent, with [O/Fe] and [Mn/Fe] having the most similar standard deviation. 

In summary, the comparison of distributions for given (\feh, \mgfe) reveals that some of the chemical cells have non-Gaussian distributions. Furthermore, some cells show a  bi-modality (e.g. the distribution of [O/Fe] in the inner disk in cell (2,2) in Figure \ref{fig:violins2}). To compare the non-Gaussian \xfe\ qualities of the radial regions, we investigate the difference in skew and kurtosis (third and fourth rows of Figure \ref{fig:Moments}). Overall, the skew and kurtosis are typically biased (non-zero differences; negative for some elements and positive for others). This is indicative of systematic trends in these moments across the [Fe/H]--[Mg/Fe] cells. The conditional element abundance Al shows a particularly large bias in overall skew across the [Fe/H]--[Mg/Fe] cells in the solar neighbourhood versus outer disk. 

\subsection{III: Overall scatter and bias in conditional abundances across Galactic radii} 
\label{sec:results III}

We have so far empirically established that the abundance gradients and \xfe\ distributions vary for different stars of a given \mgfe\ and \feh. In this section, we seek to quantify and summarise these differences, by directly comparing the results of the mean \xfe\ values of each chemical cell in the inner disk, solar neighborhood, and outer disk. To do this, we use two metrics, intrinsic dispersion and bias of conditional abundances between radial bins. Intrinsic dispersion (\sigint) tells us what the overall scatter in each [X/Fe] abundance is, between the radial bins, across the (\feh, \mgfe) cells, after accounting for measurement uncertainty. We are calculating the intrinsic dispersion around the 1:1 line for data that has been generated from groups of stars - using our (\feh, \mgfe) cells. Therefore, the confidence on each of these data points is very high, in both the x- (inner radial region) and y- (outer radial region) directions (typical standard errors on the mean are $\leq$ 0.01 dex). The scatter around the 1:1 line is therefore almost entirely driven by the intrinsic scatter in the abundances between the two radial bins. However, as an estimate to account for the additional scatter caused by the uncertainty on the mean data points, we define the intrinsic scatter as \sigint $= \sqrt{RMS - \sigma^2_\text{measurement}}$, where $RMS$ is the root mean square of the chemical cells' mean abundances about the 1:1 line, and the estimated $\sigma_\text{measurement}$ is the average standard error around the mean \xfe\ measurement for each data point. We average across all (\feh, \mgfe) cells for the radial regions considered to obtain this.

The amplitude of the intrinsic scatter is a measure of how much the overall abundance across the full set of chemical cells is non-identical, and not fully predicted by (\feh, \mgfe), between the radial bins used. Bias, on the other hand, describes the systematic difference between the two radial regions, defined as the mean difference between the mean abundances of \feh--\mgfe\ cells for two radial regions. A positive bias indicates that the inner region has systematically higher \xfe\ for given (\feh, \mgfe). The bias is effectively a measure of the overall mean conditional abundance gradient, as it is calculated here using all of the chemical cells together. These metrics serve as a useful basis for comparison to other literature. 

\begin{figure*}
     \centering
     \includegraphics[width=.9\textwidth]{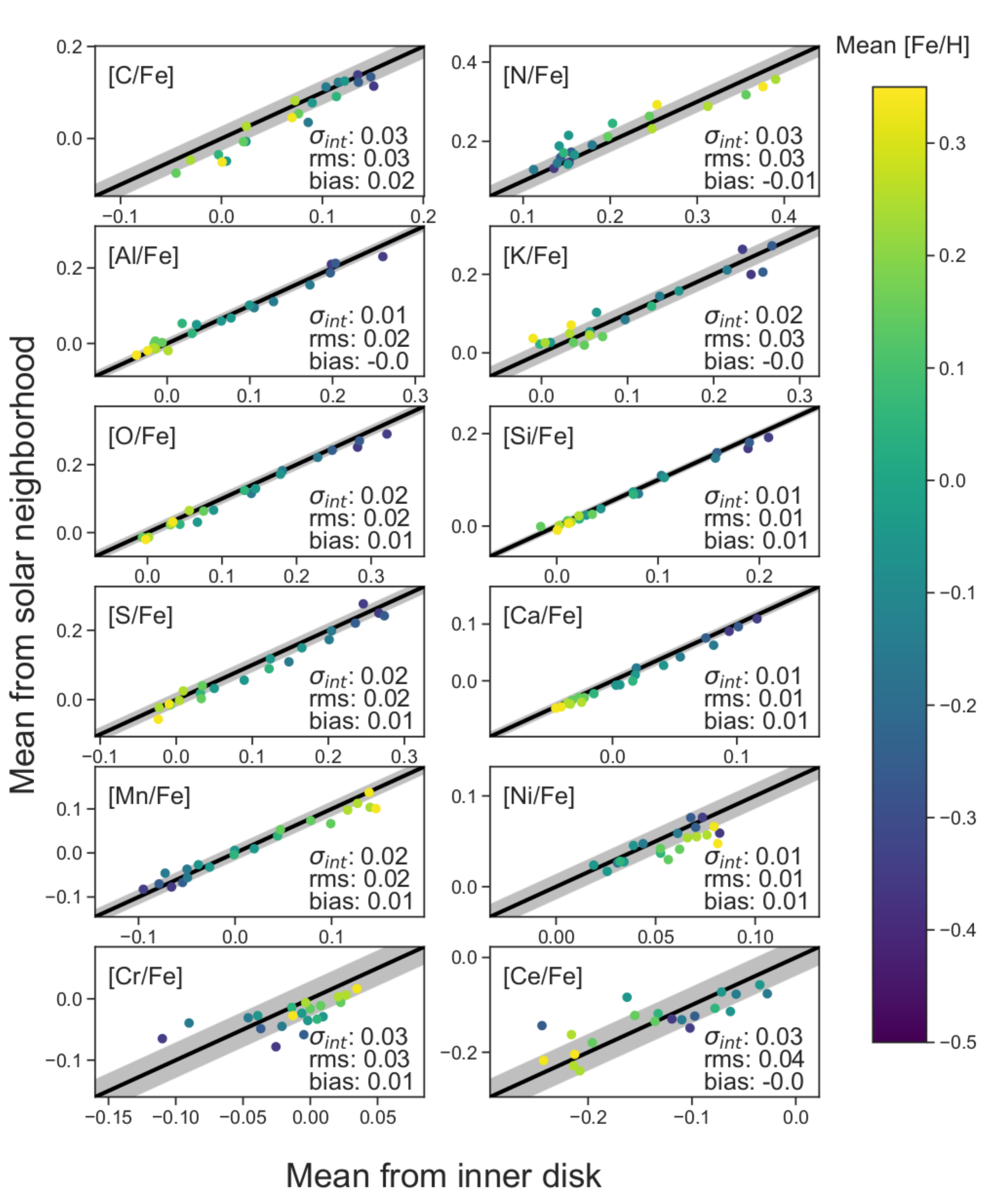}
\caption{Each point corresponds to a \feh--\mgfe\ chemical cell. The x and y values of the point represent the mean \xfe\ (measured by \apogee) of the bin in the inner disk (\rgal\ = $3-5$ kpc) and the solar neighborhood (\rgal\ = $7-9$ kpc) respectively, with the color corresponding to the mean \feh\ of the cell. The 1:1 line is colored black, with the grey bar width representing the RMS. The RMS, \sigint, and bias are given in the bottom right corner of each \xfe\ panel.}
\label{fig:lucy_means_InnVsSol}
\end{figure*}

\begin{figure*}
     \centering
     \includegraphics[width=.9\textwidth]{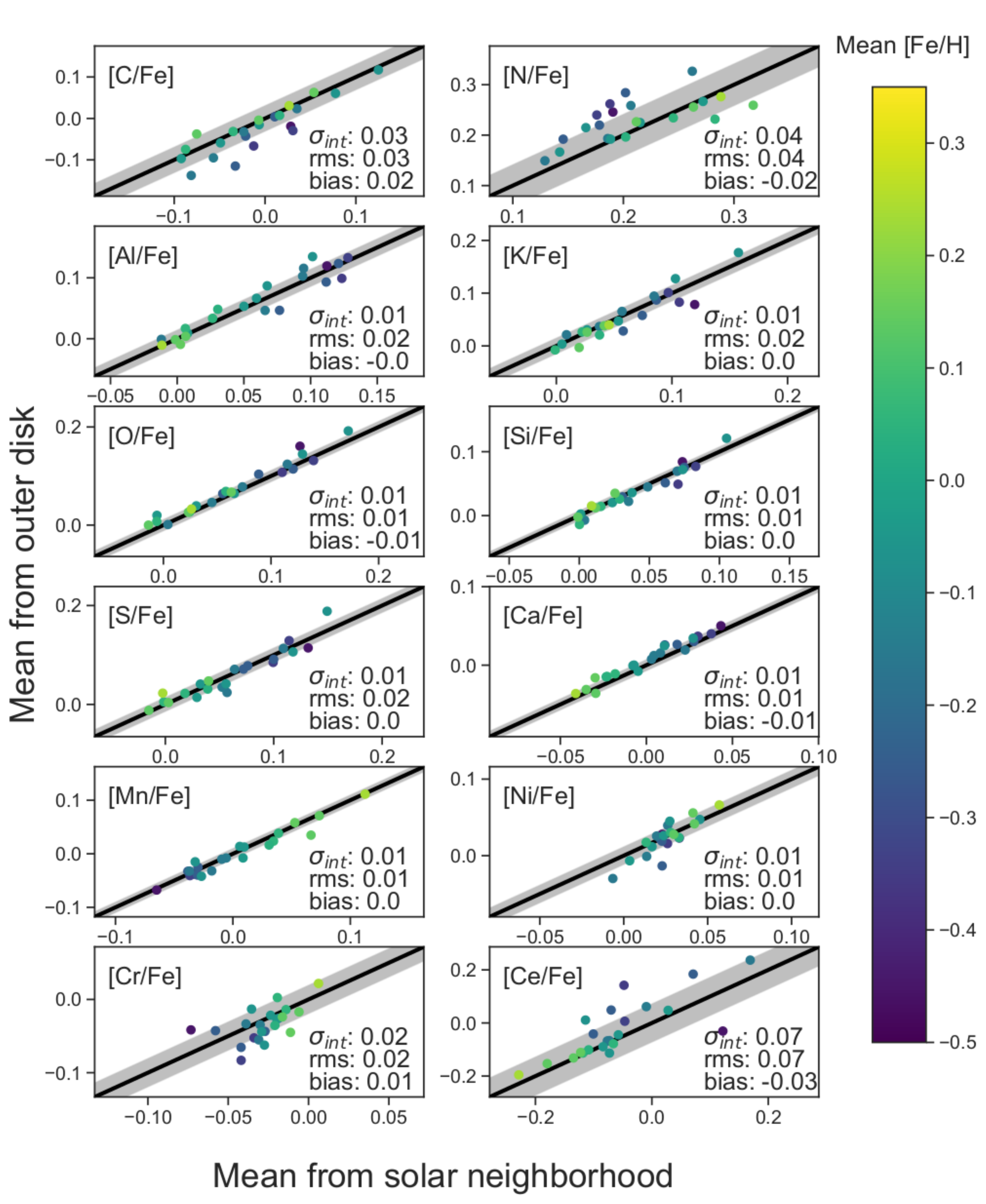}
\caption{Same as Figure \ref{fig:lucy_means_InnVsSol} for the solar disk (x-axis) and outer disk (y-axis).}
\label{fig:lucy_means_SolVsOut}
\end{figure*}

Figure \ref{fig:lucy_means_InnVsSol} shows the mean \xfe\ value for the inner disk (x-value) and solar neighborhood (y-value) for each chemical cell, with the bias, \sigint, and $RMS$ given in the lower right hand corner of each panel. Similar to Figure \ref{fig:lucy_means_InnVsSol}, Figure \ref{fig:lucy_means_SolVsOut} shows the mean \xfe\ value for the solar neighborhood (x-value) and outer disk (y-value) for each \feh--\mgfe\ cell. These figures show how some \xfe\ abundances change across Galactic radii for given (\feh, \mgfe) as a function of \feh\ (e.g. the low \feh\ chemical cells on average have higher [Ce/Fe] in the outer disk than in the solar neighborhood), and as a function of \xfe\ (e.g. chemical cells with lower \cfe\ have even lower abundance in the solar neighborhood than in the inner disk).

Looking at different nucleosynthetic families, the mean \xfe\ values for the different chemical cells fall on or near the 1:1 line for the $\alpha$-elements (O, Si, S, Ca). With a mean \sigint\ of 0.01 dex and mean absolute bias of 0.01 dex, the $\alpha$-family is on average the most similar throughout the Milky Way disk for a given value of \mgfe\ and \feh. Despite their numerically small differences, it is interesting to report that the $\alpha$- elements shown consistently have lower conditional \xfe\ abundance in the solar neighborhood than in the inner disk. O and Ca also show higher conditional abundances in the outer disk than solar neighborhood, though again these traces are minute ($\approx$ 0.01 dex).

The iron peak (Mn, Ni, Cr) and light elements with odd Z (Al, K) also show small differences across Galactic radii, with mean \sigint\ and absolute bias of 0.02 dex and 0.01 dex respectively. However, Figure \ref{fig:lucy_means_InnVsSol} shows that the majority of \feh--\mgfe\ chemical cells show a negative gradient in [Cr/Fe], with two low [Cr/Fe] cells driving the bias to be smaller. 

The light elements (C, N) and s-process element (Ce) show the largest change in \xfe\ as a function of \rgal. The average \sigint\ and absolute bias for the light elements is 0.03 dex and 0.02 dex respectively, while the averages for Ce are 0.05 dex and 0.02 dex. \cfe\ and \cefe's higher bias throughout the disk is consistent with the slopes examined in Section \ref{sec:results I}. Even though \cefe\ has the largest measurement uncertainty (0.08 dex), the scatter seen when comparing the inner disk to the solar neighborhood to the outer disk cannot be explained by measurement uncertainty alone. This perhaps shows the power of the s-process to inform Milky Way evolution studies, even at fixed (\feh, \mgfe).

\subsubsection{Additionally conditioning on age}
\label{sec:results_apogeeAge}

\begin{figure*}
     \centering
     \includegraphics[width=.825\textwidth]{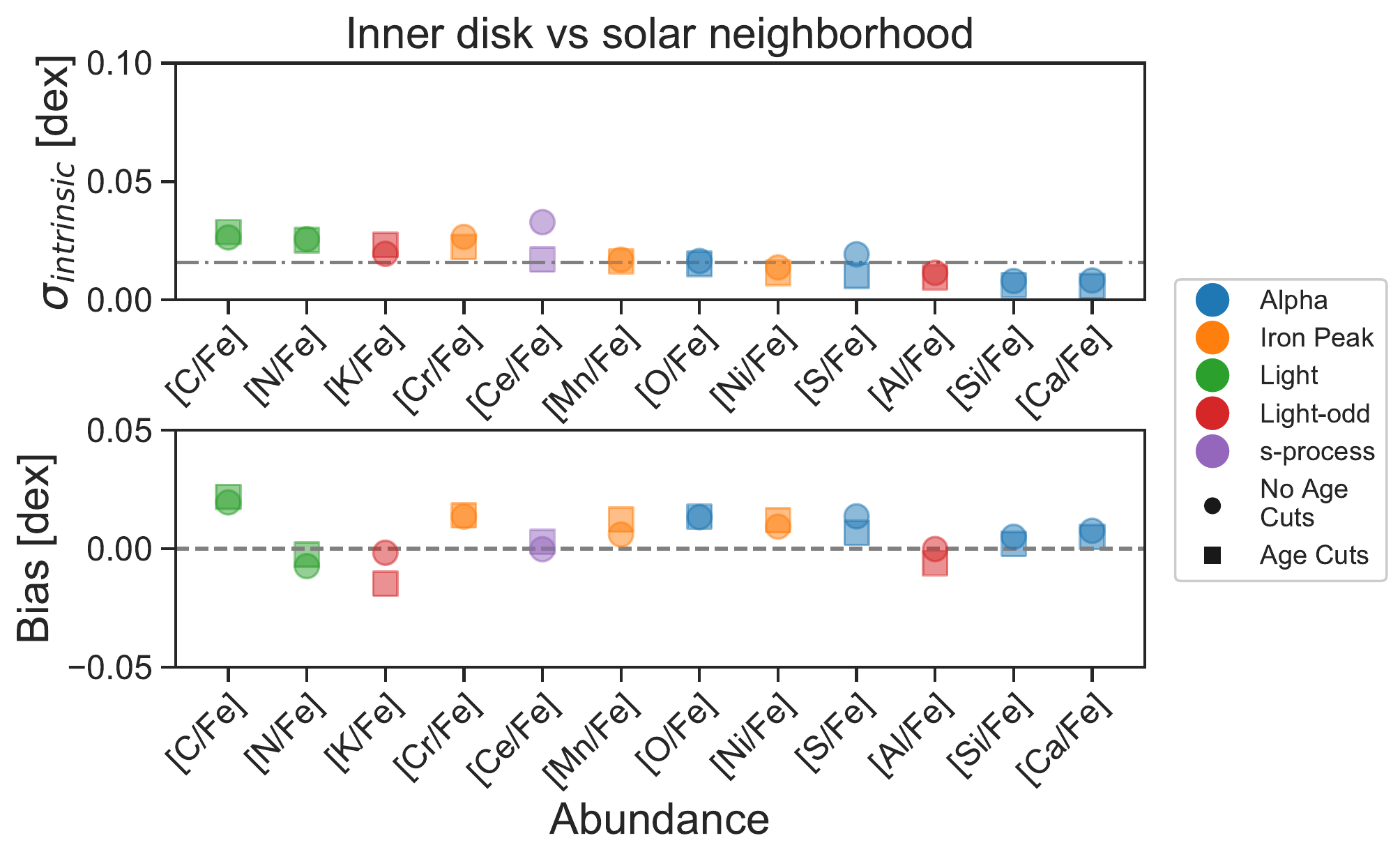}\\
     \includegraphics[width=.825\textwidth]{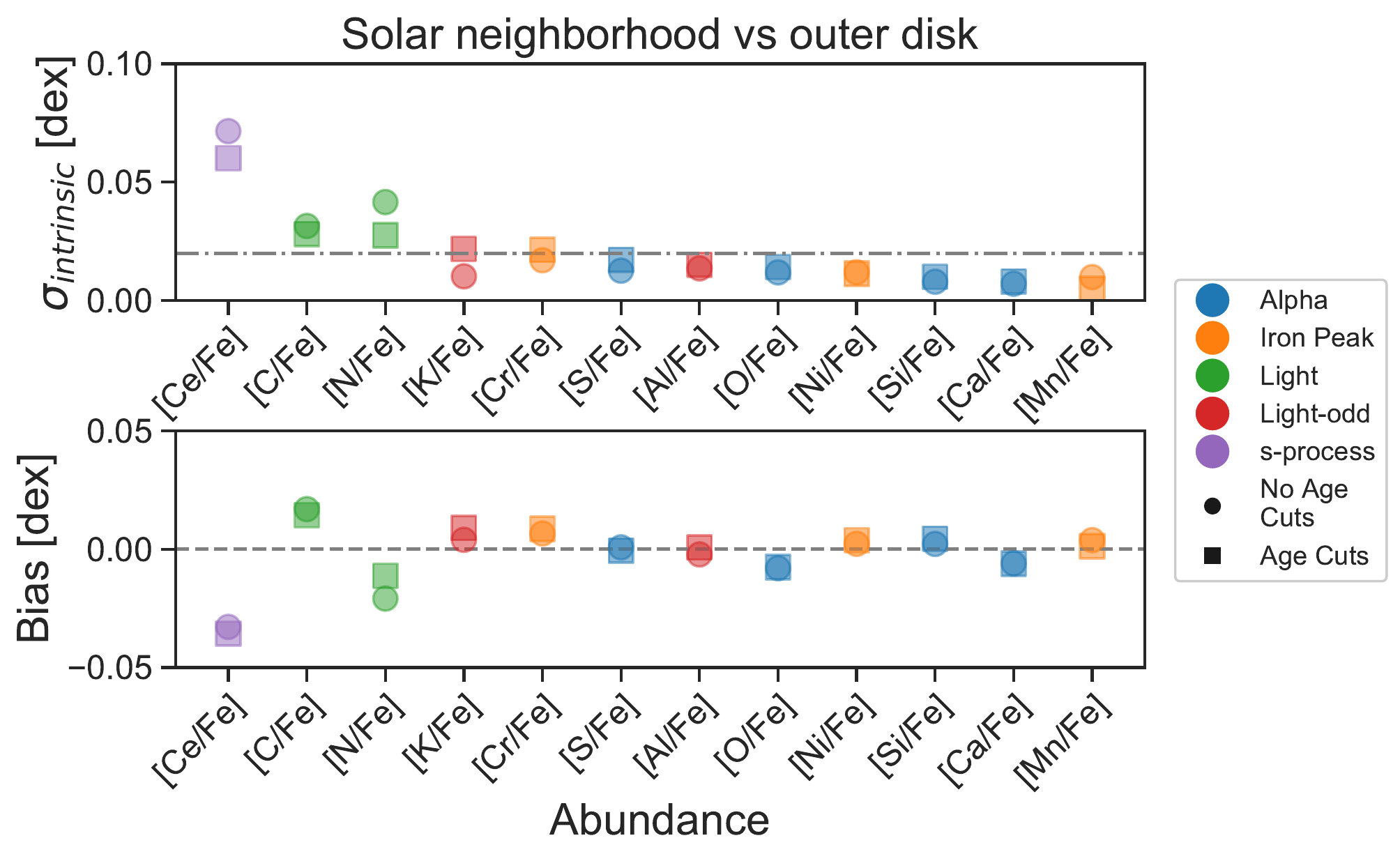}\\
\caption{The intrinsic dispersion and bias for each element colored according to their nucleosynthetic family for the inner disk vs solar neighborhood (top two panels) and solar neighborhood versus the outer disk (bottom two panels). The circular points are the bias and \sigint\ for the \feh--\mgfe\ chemical cells, and the square points are when additionally conditioning on age. The mean \sigint\ (shown with a dash-dotted line) is 0.02 dex for both comparisons. \cefe\ has the most significant difference as a function of galactic radius for fixed [Mg/Fe], [Fe/H], and age. Determined using 100 monte carlo samples, the uncertainty in \sigint\ and bias is $<0.005$ dex for the solar neighborhood and outer disk comparison. The uncertainty in \sigint\ and bias is $<0.01$ dex for the inner disk and solar neighborhood comparison, with all elements except [Ce/Fe] having uncertainties $<0.005$ dex. }
\label{fig:lucyAgeSep_biasIntr}
\end{figure*}

So far in Section \ref{sec:results III}, we have only examined the scatter and systematic differences between radial regions conditioned on supernovae contribution. Section \ref{sec:slopes_age} revealed that the steepness of radial abundance gradients depends on stellar age. Therefore, we now additionally consider stellar age in our analysis, using age bins of $0-3$ Gyr, $3-6$ Gyr, $6-9$ Gyr, and 9+ Gyr. Figures \ref{fig:lucyAgeSep_biasIntr} and \ref{fig:galah_bias} summarizes the intrinsic scatter and the bias for the elements in \apogee\ and \galah\ respectively. For these figures, we show \sigint\ and bias with (square points) and without (circular points) age as a conditional label.

Overall, the bias and \sigint\ are only marginally affected by additionally conditioning on stellar age. The mean intrinsic scatter stays at 0.02 dex and the median absolute bias stays below 0.01 dex. The only nucleosynthetic families affected by conditioning on age (i.e. their bias and \sigint\ changed by 0.01 dex) are light, light odd-z, and the s-process element cerium. Figure \ref{fig:lucyAgeSep_biasIntr} shows that most notably, conditioning on age decreases the intrinsic scatter for the elements Ce, C, and N when comparing the solar neighborhood to the outer disk by about 0.01 dex. Meanwhile, conditioning on age increases the \sigint\ in [K/Fe] throughout the outer parts of the disk from 0.01 dex to 0.02 dex. The absolute bias in [K/Fe] also increased from $\approx$ 0 dex to 0.01 dex. 

Figure \ref{fig:lucyAgeSep_biasIntr} shows that the neutron-capture element Ce shows the largest amplitude of intrinsic scatter and bias of the \apogee\ elements at 3 and 5 times above the median values, in particular across the outer areas of the disk. The large negative bias in \cefe\ between the solar neighborhood and outer disk (where there are more low-$\alpha$ stars) indicates that on average over the \feh--\mgfe\ chemical cells, \cefe\ has a positive radial gradient, agreeing with the analysis done in Section \ref{sec:results I}. 

The light-elements vary throughout the disk, with a mean \sigint\ of 0.03 dex and somewhat larger absolute biases of 0.01 dex when additionally conditioning on age. Despite the strong negative correlation between [C/Fe] and [N/Fe] \citep[see Figure 4 in][]{2020Ratcliffe}, the bias and \sigint\ in [N/Fe] shrinks towards 0 by 0.01 dex when additionally conditioning on stellar age for the solar neighborhood and outer disk analysis, while the bias and dispersion in [C/Fe] isn't affected. 
 
Overall, the light odd-z elements also show little change throughout Galactic radii when conditioning on time of birth, with a mean \sigint\ of 0.02 dex and absolute bias of 0.01 dex. As discussed previously, these differences are seen most significantly for the element [K/Fe] in this family, where conditioning on stellar age appears to highlight the unique differences, especially in the inner parts of the Galaxy. 

On the other hand, the $\alpha$- and iron peak elements show consistently low \sigint\ and bias throughout the disk for a given age, \mgfe, and \feh, showing little to no dispersion (0.01 dex and 0.02 dex, respectively) or bias (0.01 dex and 0.01 dex, respectively). 

\subsubsection{Consistency across surveys}
\label{sec:results III galah}

\begin{figure*}
     \centering
     \includegraphics[width=.9\textwidth]{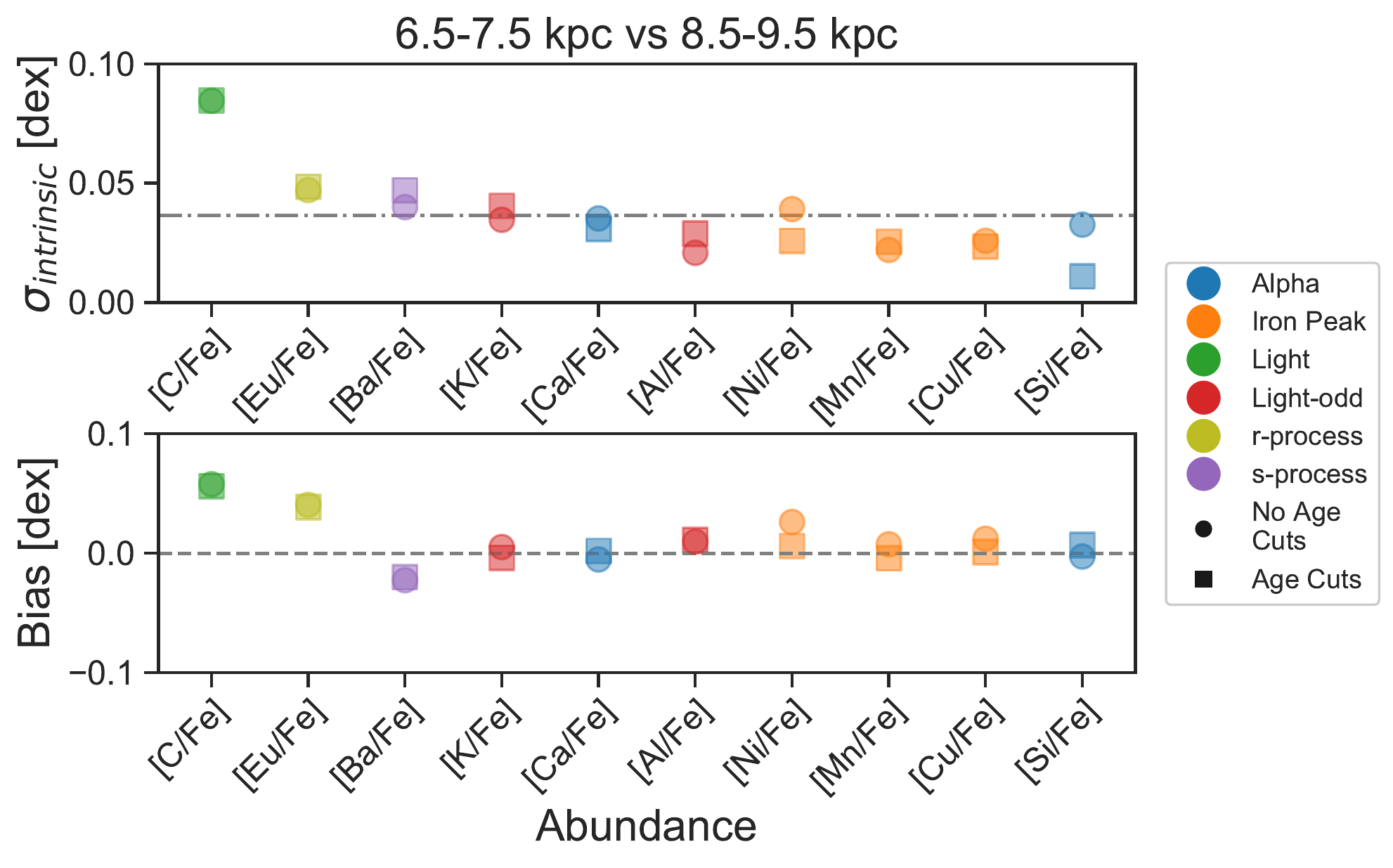}
\caption{\sigint\ and bias for \galah\ measured elements for two different radial regions in the solar neighborhood, colored according to the element's nucleosynthetic family. \sigint\ and bias are calculated using bins of (\feh, \mgfe, age). The average \sigint\ is 0.04 dex. \cfe, \eufe, and \bafe\ differ the most within the solar neighborhood. The uncertainties in \sigint\ and bias are $<0.01$ dex.}
\label{fig:galah_bias}
\end{figure*}

Figure \ref{fig:galah_bias} shows the \sigint\ and bias for the \galah\ elements in the solar neighborhood (\rgal\ = $6.5-7.5$ kpc vs \rgal\ =  $8.5-9.5$ kpc) for the chemical cells conditioned on stellar age. We find that the \sigint\ and absolute bias of the \galah\ elements are on average 0.04 dex and 0.01 dex respectively. Overall, the \galah\ abundances show consistent trends with the analysis using \apogee\ abundances in Section \ref{sec:results_apogeeAge}, with the order of the nucleosynthetic families providing additional information beyond \feh\ and \mgfe\ being the same. 

Again, the $\alpha$- (Ca, Si) and iron peak (Cu, Ni, Mn) elements show the smallest mean \sigint\ and absolute bias across the nucleosynthetic families, with mean \sigint\ of 0.02 dex and absolute bias of 0 dex. Unlike the \apogee\ analysis however, the $\alpha$-family is affected when additionally conditioning on age, with the mean \sigint\ decreasing by 0.01 dex. The light odd-z elements (K, Al) show the next smallest differences across the solar neighborhood, with mean \sigint\ and absolute bias of 0.02 dex and 0.01 dex. Consistent with the analysis done in Section \ref{sec:results_apogeeAge}, intrinsic dispersion in [K/Fe] increases when additionally using stellar age as a conditioning label, this time only marginally ($\approx0.01$ dex).

The light (C) and neutron capture (Eu, Ba) elements show the largest variation from the inner solar neighborhood (\rgal\ = $6.5-7.5$ kpc) and outer solar neighborhood (\rgal\ =  $8.5-9.5$ kpc). The bias and \sigint\ in [C/Fe] is significantly higher than that found in Section \ref{sec:results III}, with values 0.06 dex and 0.08 dex respectively. This difference could possibly be due to the different evolutionary state of the \galah\ main sequence versus \apogee\ red clump stars. The second largest differences are seen for \eufe, which has both a high \sigint\ (0.05 dex) and bias (0.04 dex), similar to \cefe's difference in the outer regions of the disk. In fact, almost every single \feh--\mgfe--age cell shows a larger \eufe\ abundance in the inner parts of the solar neighborhood compared to the outer parts of the solar neighborhood (see Figure \ref{fig:galah_ageGroupMeans} in the Appendix). \bafe\ has a bias and \sigint\ of -0.02 dex and 0.05 dex,  respectively, showing similarities to \cefe. The bias in \bafe\ is smaller than that of \cefe, potentially speaking to the overall production of Ba versus Ce in Asymptotic Giant Branch (AGB) stars.

\section{Discussion}\label{sec:discussion}

This work is an introductory exploratory analysis of chemical abundance trends across the Milky Way disk. It provides an assessment of individual element information across radius at fixed (\feh, \mgfe) and serves to model analysis possibilities when the sample size of survey data increases, for the same or higher precision measurements.  Overall individual element abundance gradients for the field \citep[e.g.][]{Eilers2022} and open clusters \citep[e.g.][]{2022spina} are important to understand nucleosynthesis as a whole and improve chemical evolution models. However, we can add strong additional model constraints, and advance our understanding of galaxy evolution, by examining the conditional abundances over radius. Furthermore, we can learn which elements are most useful to work with to get additional discriminating power of birth environments at fixed (\feh, \mgfe) and also fixed (\feh, \mgfe, age). Our key findings are summarized below: 

\begin{enumerate}
    \item \xfe\ abundance gradients vary for different \feh--\mgfe\ populations with some \rgal--\xfe\ relations being complex and nonlinear (Figure \ref{fig:runningMean}). While it is expected that the iron peak and $\alpha$-elements would show small trends in \rgal\ since we are conditioning on their sources of enrichment, the absolute value of most slopes, including elements from all nucleosynthetic families, is less than 0.03 dex/kpc (Figure \ref{fig:apogee_slopes}).
    
    \item The strength of the \rgal--\xfe\ relationship changes as a function of age for some abundances. Younger stars in the low-$\alpha$ sequence have steeper slopes compared to older stars (Figure \ref{fig:cefe-slope}).
    
    \item The distributions of \xfe\ vary as a function of \feh--\mgfe\ chemical cell and radial location in the Galaxy (Figures \ref{fig:violins2} and \ref{fig:violins1}).  
    
    \item There is a mean intrinsic scatter and absolute bias across the Milky Way disk for fixed (\feh, \mgfe) of 0.02 dex and 0.01 dex respectively, for \apogee\ measured abundances (\rgal\ = $3-13$ kpc). The mean intrinsic scatter and bias is 0.04 dex and 0.01 dex, respectively, for \galah\ abundances for fixed (\feh, \mgfe) (\rgal\ = $6.5-9.5$ kpc). Additionally, conditioning on stellar age only marginally affects these mean measurements (Figures \ref{fig:lucyAgeSep_biasIntr} and \ref{fig:galah_bias}). Seven \apogee\ elements (corresponding primarily to the $\alpha$ and iron peak) have intrinsic dispersions under 0.02 dex and small biases ($<0.015$ dex) (Si, Ca, Ni, S, Al, O, Mn), indicative they have a very marginal additional birth information beyond (\feh, \mgfe, age). The scatter and bias found in across all elements are robust to different chemical cell widths chosen. 
\end{enumerate}

In addition to the overall results given above, we find trends as a function of nucleosynthetic family. Unless specifically mentioned otherwise, the below results are for abundances measured by \apogee.

The \textbf{$\alpha$-} (S, O, Si, Ca) and \textbf{iron peak} (Cr, Mn, Ni) abundances have small radial gradients for given \feh\ and \mgfe, ranging between $-0.01$ dex/kpc and 0.01 dex/kpc, with S and Cr showing the steepest slopes. The representative $\alpha$ and iron peak elements ([Mn/Fe] and [O/Fe]) overall show the smallest variation in distributions for given \feh\ and \mgfe\ compared to the other representative abundances, with absolute median differences in mean and standard deviation of 0.01 dex, and absolute differences in skew and kurtosis of $\approx$ 1.5 dex. Of the $\alpha$- and iron peak elements looked at in this work measured by \apogee, all except Cr showed minimal bias ($<0.015$ dex) and \sigint\ ($<0.02$ dex). The bias and \sigint\ in [Cr/Fe] is 0.01 dex and $\approx0.02$ dex.

\textbf{Light with odd Z} abundances (Al, K) show small radial gradients for given (\feh, \mgfe) (0.01 dex/kpc). Comparing the distributions of each chemical cell across the different radial regions, [Al/Fe] showed minimal difference in mean abundance ($\sim$0 dex) and scatter (0.01 dex) across the disk. However, [Al/Fe] showed the largest differences at higher moments, with a difference in skew and kurtosis of 0.9 dex and 3 dex. While [Al/Fe] had one of the smaller biases ($\sim$ 0 dex) and \sigint\ ($<$ 0.02 dex) across the disk, [K/Fe] showed above average \sigint\ (0.03 dex) and bias (0.01 dex) that increased when additionally conditioning on stellar age. 

The \textbf{light} elements (C, N) have conditional radial gradients up to 0.02 dex/kpc. The negative gradient in \cfe\ is seen in most \feh--\mgfe\ chemical cells, with the difference in mean abundance between the different radial regions showing a bimodality. The differences in the higher statistical moments of \cfe\ for the different chemical cells are centered around 0 dex, except for the different in kurtosis between the inner disk and solar neighborhood, where the solar neighborhood has higher kurtosis values. The light elements that evolve over time show scatter---that does not disappear with age---which is presumably explained by evolutionary changes.

The \textbf{neutron capture} elements (Ce, Ba, Eu) showed the largest variation across the disk. Conditional \cefe\ slopes range between -0.01 dex/kpc and 0.03 dex/kpc, and the first two statistical moments show the largest variation in \cefe, in particular in comparing the solar neighborhood and outer disk. The large intrinsic scatter of \cefe, \bafe\ (\galah), and \eufe\ (\galah) (all $\approx0.05$ dex) illustrates the diversity of the s- and r-processes for a given supernovae contribution conditioned on stellar age (Figures \ref{fig:lucyAgeSep_biasIntr} and \ref{fig:galah_bias}). 

Our findings on the radial gradients in \xfe, in particular Figure \ref{fig:apogee_slopes}, speak to the different formation histories of the high- and low-$\alpha$ sequences. We find that the \rgal--\xfe\ gradients in the high-$\alpha$ sequence show no correlation with \feh. That is, we see no evidence for any initial gradients in the element abundances with [Fe/H] that have been preserved to the present-day. Furthermore, the majority of the high-$\alpha$ chemical cells show near-zero abundance gradients. This is resonant with \cite{2015Haywood}, who proposed that the thick disk formed out of homogeneous gas. However, we find that there are some elements such as [Ce/Fe] that show some chemical cells in the high-$\alpha$ sequence reaching conditional abundance gradients of $\approx$ $\pm$0.02 dex/kpc. While \cite{2021A&A...649A.126T} found the [Ce/Fe] gradient in the thick disk to be negligible (-0.002 dex/kpc), our work shows the additional information available by considering the Milky Way's chemical evolution through a division into chemical cells. 

Similar to other works on the $\alpha$-sequences as a whole \citep[e.g.][]{2014A&A...572A..33M, 2014A&A...567A...5R}, we find that the low-$\alpha$ sequence shows the largest conditional abundance gradients for the non-$\alpha$, and non-iron peak elements, with exception in [Al/Fe]. Furthermore, by examining individual chemical cell populations, we find that many abundances (C, N, S, Ca, Mn, Ce) in the low-$\alpha$ sequence show systematic conditional gradient trends as a function of \feh. This structure is presumably inherited from initial abundance gradients at birth, similarly to the initial [Fe/H] radial abundance gradient \citep[e.g.][]{2018Minchev_rbirth}. The correlation in the gradient trends in the low-$\alpha$ sequence with \feh\ was likely also even stronger in the past. Importantly, we uncover in Figure \ref{fig:apogee_slopes} that the most significant non-zero conditional abundance gradients in the low-$\alpha$ sequence are seen for stars with [Fe/H] $< -0.2$ dex. This can also be seen in the simulation modeling and observational data in \cite{2021Johnson}, who show the mode of [O/Fe] decreases faster in \rgal\ for stellar groups with \feh\ $< -0.2$ dex (see Figure 12 in their work). For all chemical cells with [Fe/H] $> -0.2$ dex, across all element families, we find the absolute gradients in chemical cells range from 0 -- 0.01 dex/kpc. In particular, \feh--\mgfe\ chemical populations with \feh\ $>$ 0 dex show absolute gradients $\sim$0 dex/kpc for nearly all elements, with only C, N, K, and Ce showing slightly larger trends ($<0.01$ dex/kpc).

In order to quantify our results and to compare to prior work, we use two metrics to understand an element's ability to link back to their birth environment: intrinsic dispersion (\sigint) and bias. \sigint\ informs us if the differences in the abundances at different radii at fixed (\feh, \mgfe, age) can be explained by the measurement uncertainty and quantify the level to which (\feh, \mgfe, age) are not fully predictive of the abundances. The bias is an integrated measurement of the radial conditional abundance gradients. We report a mean intrinsic dispersion and absolute bias of 0.02 dex and 0.01 dex, respectively, for \apogee\ measured abundances across the disk (\rgal\ = $3-13$ kpc). We report a mean intrinsic dispersion and absolute bias of 0.04 dex and 0.01 dex, respectively, for \galah\ abundances across the solar neighborhood (\rgal\ = $6.5-9.5$ kpc). The neutron capture elements show the largest variation overall, with \sigint\ $\approx$ 0.05 dex and a mean absolute bias of 0.03 dex.

The level of chemical homogeneity for fixed \feh, \mgfe, and age in the current day disk that we find is in agreement with previous work. With a two-process model that uses the weighted sum of SNII and SNIa, \cite{2021Weinberg}, \cite{2021Griffith}, and \cite{2021Griffith_residuals} showed that residuals for most abundances in the Milky Way disk are similar for the 16 elements they studied ($\leq 0.07$ dex), with the residual in Ce being significantly larger ($\sim$ 0.1 dex). \cite{2022Ness} goes further to argue that the disk is homogeneous in supernovae elements to 0.01 -- 0.015 dex at fixed birth radius and time. By conditioning on \feh, \mgfe, \teff, and \logg, \cite{2021TingWeinberg} find the residual scatter for most [X/H] abundances reconstructed using the machine learning technique normalizing flow to be $\leq$ 0.02 dex. This scatter, measured for an integrated population across across the entire disk (\rgal\ = $3-13$ kpc), is comparable to the median \sigint\ that we find in our work between radial bins. This similar dispersion possibly speaks to the role of radial mixing in the Milky Way, showing that at any given radius, stars are from a range of stellar birth populations. \cite{2021TingWeinberg} also indicate that many elements have additional independent information beyond supernovae elements, which is a similar conclusion that we draw when additionally conditioning on age and Galactic radius in a model-free way. However, we emphasise that the amplitude of this additional information is very small. 

Under radial migration, we might expect stars at any given location in the disk to be drawn, with some varying density, from the same underlying chemical distribution. The biases in our conditional abundances are indicative that there are marginal differences in the underlying abundance distributions at fixed (\feh, \mgfe) and also age. The conditional gradients however show that these biases originate only from some regions in the (\feh, \mgfe) plane (not all cells show gradients in each element). A non-zero \sigint\ between radial bins also indicates some difference, but one that is not necessarily systematic, and so a product of either abundance scatter at birth, or evolutionary impacts. 

The bias (and \sigint) is most significant for elements Ce, Eu, C, and Ba, and means they must have additional power in resolving birth environment, and radius in the Milky Way beyond (\feh, \mgfe). Importantly however, we note that we can see using chemical cells, that the most significant non-zero conditional gradients across radius are seen for stars with [Fe/H] $<$ --0.2 dex. At the same time, Figure \ref{fig:Moments} shows that even at high [Fe/H], the conditional distributions differ in detail in their standard deviation, skew and kurtosis. These differences, which are systematic (see Figure \ref{fig:Moments} which shows non-zero distributed moments) might be explained, in part, by initial non-Gaussian distributed element abundance distributions and radial migration, at least for the populations with near-zero radial gradients, above [Fe/H] $>$ 0 dex. We also note that when conditioning on stellar age in addition to our (\feh, \mgfe) cells, the \apogee\ sample shows consistency with radial migration expectations \citep[][]{Frankel2018}, with older stars showing weaker \rgal--\xfe\ gradients (Figure \ref{fig:cefe-slope}). 

\subsection{Physical implications}
The scatter we find for element abundances within \feh--\mgfe\ chemical cells may be caused by chemical inhomogeneity of birth clusters (from mixing or production stochasticity), mass dependence of chemical yields (e.g. SNII), fractional changes in source contribution over time \citep[][]{Kobayashi2020}, or additional sources beyond SNII and SNIa (e.g. AGB, r- and s-processes). The non-zero scatter may also suggest that stellar abundances are not time-invariant, and evolve throughout a star’s lifetime. The radial systematic changes reported in this work illustrate that radial migration has not fully mixed the stellar populations in the disk, and some \rbirth\ signatures of different star formation environments still exist across radius in the individual abundances at fixed (\feh, \mgfe), and age. These systematic changes may imply the relative proportion or mass distribution of SNII and SNIa can vary to achieve the same amount of (\feh, \mgfe). Therefore, stars within the same chemical cell could have be born in differently enriched environments. Pairing this idea with underlying environment parameters, a radially varying initial mass function \citep[see][(although changes must be small) \cite{2021Griffith}]{Horta2021,Guszejnov2019}, as well as a radially varying star formation rate and star formation efficiency \citep{2017A&A...605A..59R} may explain the conditional gradients observed.

\subsection{Additional analysis accounting for residual \xfe--\teff\ dependencies} \label{sec:temp}

Some abundances change as a function of temperature, due to both measurement systematics, stellar evolution, and atomic diffusion \citep[e.g.][]{2019Jofre,Liu2019,2017ApJ...840...99D}. Therefore we deliberately chose narrow temperature ranges for our analysis (Section \ref{sec:data}), even though this significantly reduced our stellar sample. We wish to demonstrate that our results are not significantly affected by, nor an artifact of, any marginal remaining abundance differences as a function of temperature that might persist across our narrow selection. To do this, we repeat the analysis in Section \ref{results} on  abundances where we remove any temperature dependence. Some work, such \cite{2021Weinberg}, uses linear model fitting to correct abundances for \teff\ affects, while other work \citep[e.g.][]{2021TingWeinberg, 2022Ness} conditions  a model on \teff\ and \logg\ to avoid systematic uncertainties that affect abundance measurements. Recently \cite{Eilers2022} propose a model to correct for ``nuisance" parameters such as \logg\ and \teff. Here, we fit a linear model in each \teff--\xfe\ plane and predict the \xfe\ abundance for a given \teff\ (\xfe$_\text{pred}$). We define the corrected abundance values as $$\xfe_\text{corr} = \xfe\ - (\xfe_\text{pred} - \overline{\xfe}).$$ 

Our results with the corrected \xfe\ show minor differences with the uncorrected \xfe\ done in Section \ref{results}. As a whole, the mean \sigint\ and absolute bias shrinks by less than 0.01 dex when repeating the analysis done in Section \ref{sec:results III}. The largest differences are for the light, light odd-Z, and iron peak families. For these elements, their \sigint, when comparing the inner disk and solar neighborhood, decreases by $\approx$ 0.01 dex to 0.02 dex, 0.01 dex, and 0.01 dex respectively. In particular, the bias in \cfe\ (measured by \apogee) between the inner disk and solar neighborhood also decreases from 0.02 dex to $<0.01$ dex, with \sigint\ also decreasing from 0.03 dex to 0.01 dex. Similar decreases were also seen in the investigation of \cfe\ in the solar neighborhood using \galah\ measured data. 

While most abundances show a decrease in bias and/or \sigint\ across the disk with our calibration, the \sigint\ in \cefe\ stayed consistent after adjusting for temperature effects. However, the absolute bias in \cefe\ between the inner disk and solar neighborhood increased from $\approx$0 dex to 0.02 dex. There was also a $\approx0.005$ dex increase in the \sigint\ of \eufe\ across the solar neighborhood.

Broadly, the results with our abundances calibrated for temperature are consistent, but served to reduce the overall scatter in the distributions of some elements between radial bins. These reductions are indicative that some, but not all of the fraction of scatter between radial bins is an artifact of marginal abundance bias with temperature.

\subsection{Future applications}

In the coming years, we will have access to increasing numbers, indeed tens of millions, of stellar spectra observed throughout the entire Milky Way \citep{4midable,helmi20194most,sdssV}. Applying the concepts established in this work to these surveys will provide an in depth exploration of the abundance gradients as a function of Galactic radius, azimuth, and height above the midplane, focusing on more abundances and stars than e.g. \cite{2014A&A...568A..71B,2011ApJ...738...27B}. See also \cite{2019A&A...628A..38S} for their work on azimuthal [O/H] and \feh\ abundance variations in simulations. 

One limitation of our approach in Section \ref{sec:results III} is that we focus on mean quantities for each bin, however in Section \ref{sec:results II} we showed that the \xfe\ distributions are non-Gaussian for some mono-\feh--\mgfe\ populations in different radial sections of the disk. Future work might focus on the non-Gaussianity of the conditional abundances, which may prove important. 

\section{Acknowledgements} 
M.K.N. acknowledges support from a Sloan Foundation Fellowship.

\section{Appendix - additional figures}

\begin{figure*}
     \centering
     \includegraphics[width=1\textwidth]{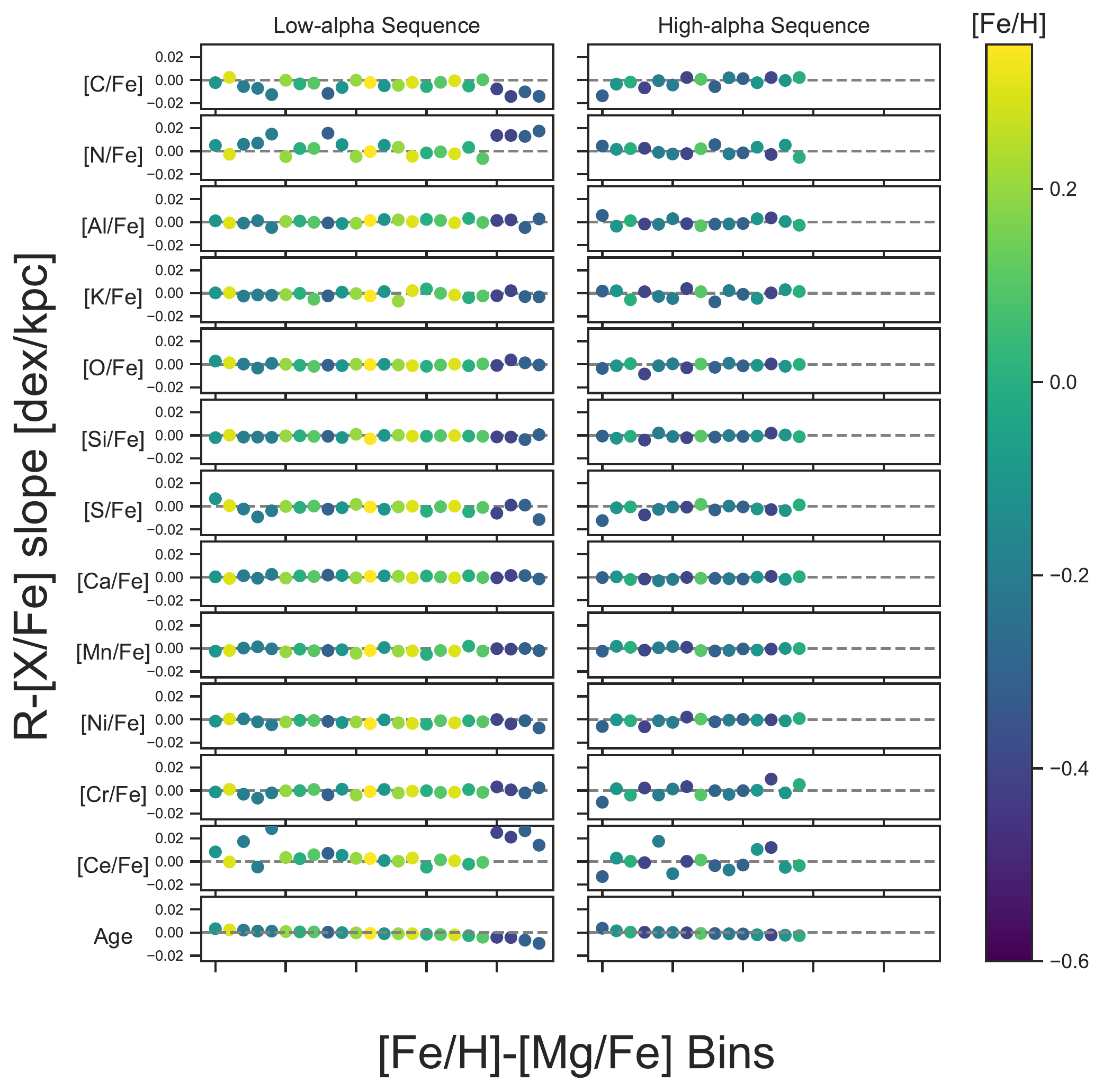}
\caption{Same as Figure \ref{fig:apogee_slopes}, with the age gradient shown in the bottom row with units of Gyr/kpc. Chemical cells are ordered by age gradient. The chemical cells with the steepest age gradients correspond to the lowest metalicity cells. The largest conditional gradients in \cefe, \cfe, and [N/Fe] correspond to the bins with the steepest age gradients.}
\label{fig:ageSlopes}
\end{figure*}

\begin{figure*}
     \centering
     \includegraphics[width=1\textwidth]{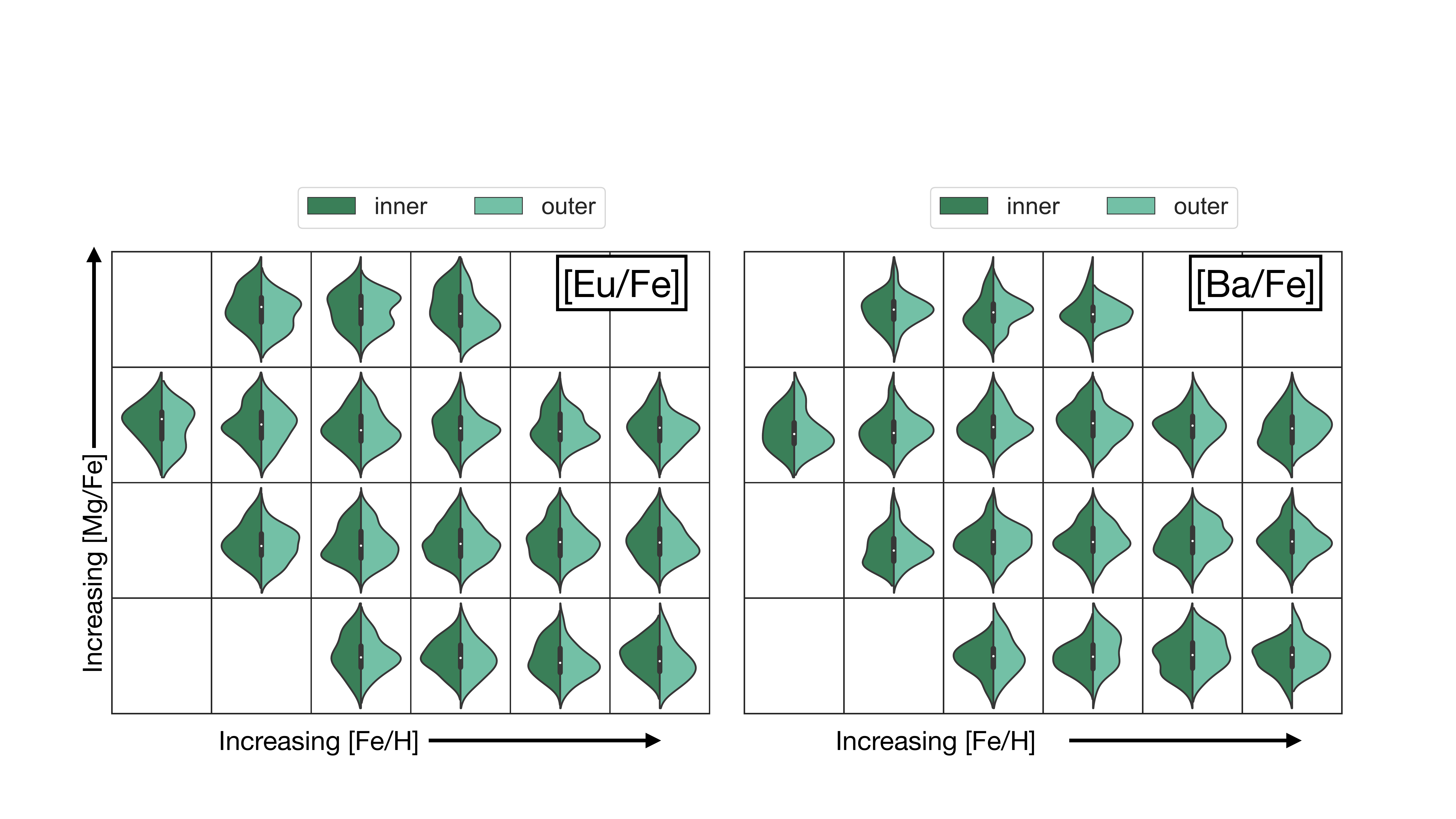}
\caption{Split violin plots showing the \xfe\ density distribution for different \feh--\mgfe\ chemical cells in the inner (6.5-7.5 kpc) and outer (8.5-9.5 kpc) solar neighborhood for \galah\ measured abundances \eufe\ and \bafe.}
\label{fig:EuBa_Violin}
\end{figure*}

\begin{figure*}
     \centering
     \includegraphics[width=.9\textwidth]{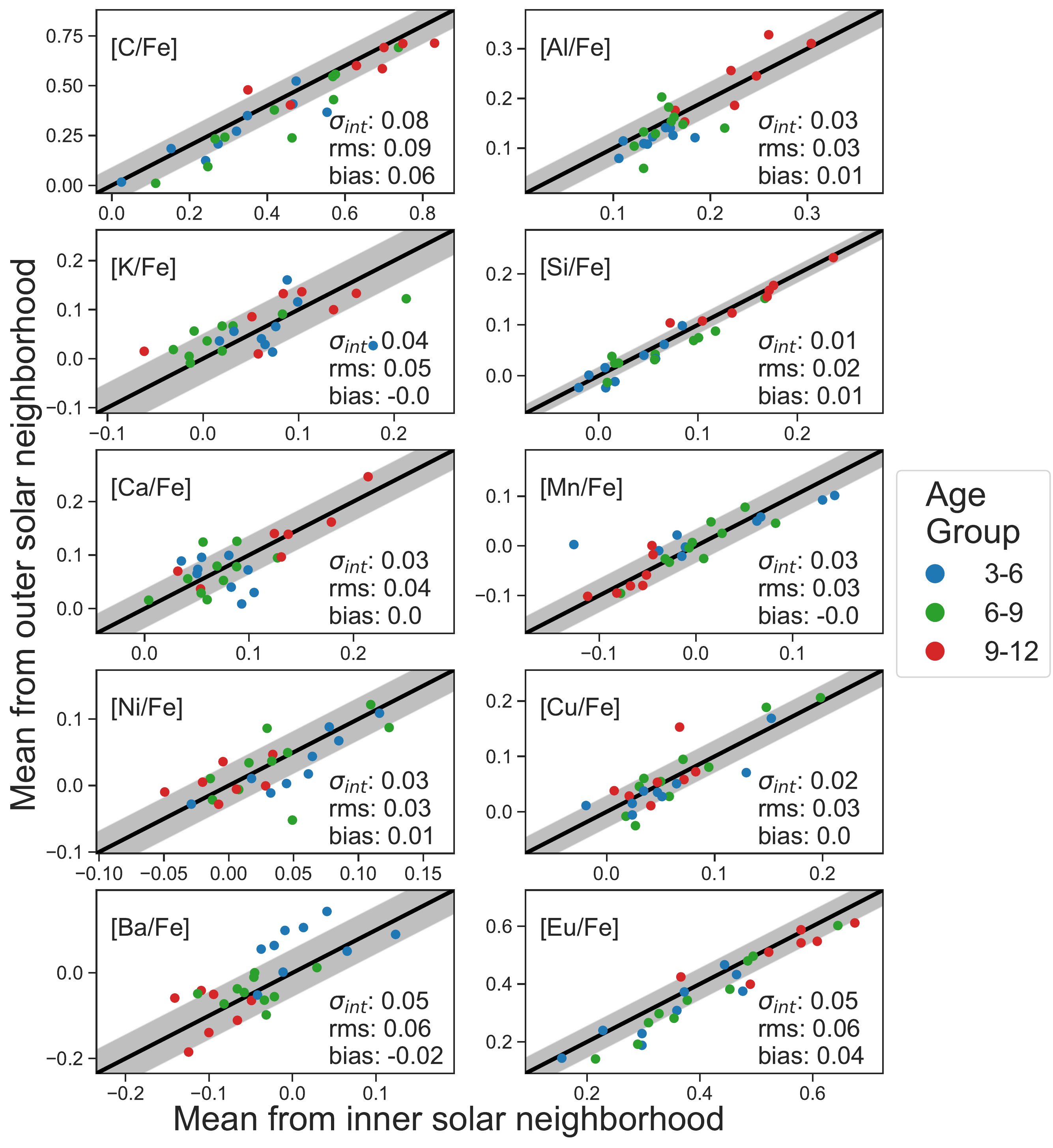}
\caption{Similar to Figures \ref{fig:lucy_means_InnVsSol} and \ref{fig:lucy_means_SolVsOut}, here we show the mean \xfe\ value for \feh--\mgfe--age cells in the inner solar neighborhood (6.5-7.5 kpc) and outer solar neighborhood(8.5-9.5 kpc) for \galah\ measured abundances. Here, we color each point by the bin's age group (3-6 Gyr, 6-9 Gyr, 9-12 Gyr). }
\label{fig:galah_ageGroupMeans}
\end{figure*}

\begin{figure*}
     \centering
     \includegraphics[width=.9\textwidth]{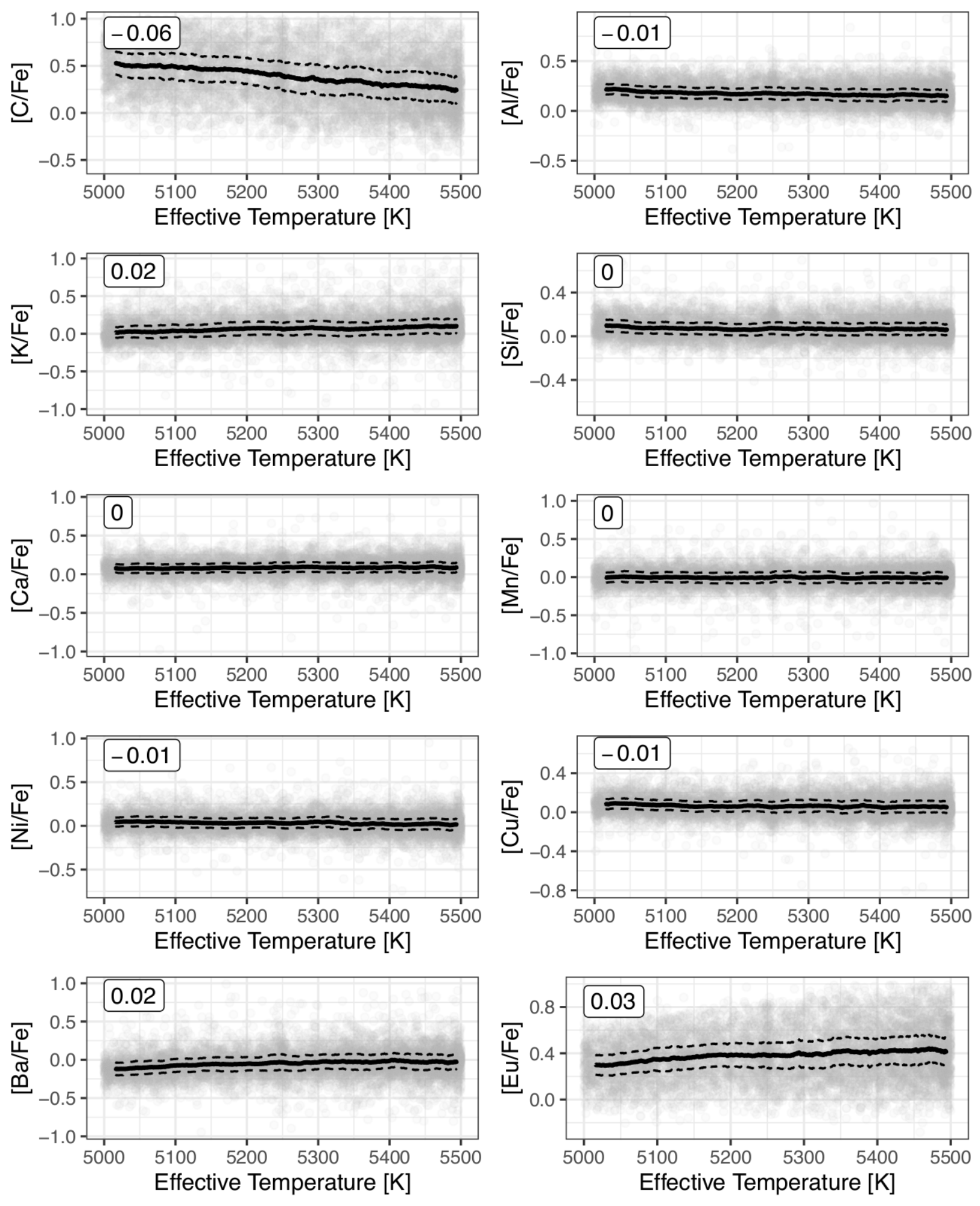}
\caption{Running mean of [X/Fe] vs effective temperature for \galah\ measured abundances. The slope per 100K is given in the top left corner of each panel. }
\label{fig:xfe-temp}
\end{figure*}

Here we include additional figures that help readers interpret results. Figure \ref{fig:ageSlopes} is very similar to Figure \ref{fig:apogee_slopes}, but with the slope of the chemical cells ordered by the cell's age gradient. Figure \ref{fig:EuBa_Violin} shows the distributions of \bafe\ and \eufe\ within different radial regions of the solar neighborhood, conditioned on \feh\ and \mgfe. Figure \ref{fig:galah_ageGroupMeans} is similar to Figure \ref{fig:lucy_means_InnVsSol}, however this figure shows the mean \xfe\ within the solar neighborhood for \galah\ measured abundances conditioned on (\feh, \mgfe, age), colored by the cell's mean age. Finally, Figure \ref{fig:xfe-temp} shows the relationship between \xfe\ and \teff\ for \galah\ measured abundances. This dependence inspired the additional testing done in Section \ref{sec:temp}.

\bibliography{AbundanceBib.bib}
\end{document}